\documentclass[fleqn,usenatbib]{mnras}

\usepackage{newtxtext,newtxmath}

\usepackage[T1]{fontenc}
\usepackage{ae,aecompl}

\usepackage{graphicx}	
\usepackage{amsmath}	
\usepackage{amssymb}	

\title[Modelling white dwarf pollution]{Modelling the distributions of white dwarf atmospheric pollution: a low Mg abundance for accreted planetesimals?}

\author[S. G. D. Turner \& M. C. Wyatt]{
Samuel G. D. Turner\thanks{E-mail: sgdt2@cam.ac.uk}
and Mark C. Wyatt
\\
Institute of Astronomy, University of Cambridge, Madingley Road, Cambridge, CB3 0HA, UK
}

\date{Accepted XXX. Received YYY; in original form ZZZ}

\pubyear{2018}

\begin{document}
\label{firstpage}
\pagerange{\pageref{firstpage}--\pageref{lastpage}}
\maketitle
\graphicspath{{Figures/}}

\begin{abstract}
The accretion of planetesimals onto white dwarf atmospheres allows determination of the composition of this polluting material. This composition is usually inferred from observed pollution levels by assuming it originated from a single body. This paper instead uses a stochastic model wherein polluting planetesimals are chosen randomly from a mass distribution, finding that the single body assumption is invalid in ${>20\%}$ of cases. Planetesimal compositions are modelled assuming parent bodies that differentiated into core, mantle and crust components. Atmospheric levels of Ca, Mg and Fe in the model are compared to a sample of 230 DZ white dwarfs for which such pollution is measured. A good fit is obtained when each planetesimal has its core, mantle and crust fractions chosen independently from logit-normal distributions which lead to average mass fractions of $f_\text{Cru}=0.15$, $f_\text{Man}=0.49$ and $f_\text{Cor}=0.36$. However, achieving this fit requires a factor 4 depletion of Mg relative to stellar material. This depletion is unlikely to originate in planetesimal formation processes, but might occur from heating while the star is on the giant branch. Alternatively the accreted material has stellar abundance, and either the inferred low Mg abundance was caused by an incorrect assumption that Mg sinks slower than Ca and Fe, or there are unmodelled biases in the observed sample. Finally, the model makes predictions for the timescale on which the observed pollutant composition varies, which should be the longer of the sinking and disc timescales, implying variability on decadal timescales for DA white dwarfs.
\end{abstract}




\section{Introduction}

The analysis of rocky planets around main sequence stars is limited to measurements of their mass and radius and thus their density (e.g. \citealt{2011Natur.470...53L}). From this, plausible compositions of the planets can be determined by comparison with models of interior planetary structure (e.g. \citealt{2010ApJ...712..974R}). This method is restricted both through the large errors usually associated with mass and radius measurements and due to the fact that the interior composition is underdetermined in this way.

A promising way to determine accurate compositions of exoplanetary material is through observations of so called polluted white dwarfs. It is now thought that as many as $50\%$ of white dwarfs exhibit metal lines in their spectra \citep{2014A&A...566A..34K}. By virtue of their high surface gravity metals in white dwarf atmospheres sink and vanish from the atmosphere on timescales of days to tens of thousands of years for stars with H dominated atmospheres and tens to millions of years for stars with He dominated atmosphere \citep{2009A&A...498..517K}. Since these timescales are much shorter than the age of the white dwarfs any metals present in white dwarf atmospheres must have arrived there recently. Many different elements have now been observed and the prevailing opinion is that these metals arrive in the atmosphere via the accretion of rocky planetesimals rather than being stellar material from a companion star or radiative levitation of deep material (\citealt{2014AREPS..42...45J}; \citealt{2014A&A...566A..34K}).

These planetesimals are relics of the planetary systems that originally formed around the main sequence stars that were the progenitors of the polluted white dwarfs. These planetesimals are then scattered into the inner system as has been shown in N-body simulations (\citealt{2011MNRAS.414..930B}; \citealt{2012ApJ...747..148D}). Scattering to within the tidal radius of the white dwarf may then cause tidal disruption of the planetesimals producing circumstellar discs which then accrete onto the white dwarf (\citealt{2014MNRAS.445.2244V}, \citeyear{2015MNRAS.451.3453V}).

With measured abundances of many different elements in the white dwarf atmospheres, their ratios can be used to determine the composition of the accreted material. This determination is complicated because different metals present in the pollutant material sink out of the atmosphere on different timescales and therefore the directly observed composition of the white dwarfs atmospheres is expected to evolve over time \citep{2009A&A...498..517K}. Nevertheless, since the first detection of polluted white dwarfs, numerous papers have analysed the composition of the pollutant material. Of these, several have concluded that the accreted material must originate from fragments of differentiated bodies (\citealt{2011ApJ...739..101Z}; \citealt{2013ApJ...766..132X}; \citealt{2015MNRAS.451.3237W}; \citealt{2016MNRAS.458..325K}; \citealt{2018MNRAS.479.3814H}).

All of the previous mentioned analyses work under the assumption that the pollution is the result of the accretion of a single object. However, \citet{2014MNRAS.439.3371W} showed that the accretion of multiple bodies are required to explain the way in which the mass of pollutants in the white dwarf atmospheres depend on the timescales for those metals to sink out of the atmosphere. Given that there is evidence for differentiated planetesimals, it is natural to consider the role of multiple accretions in how the observed composition relates to the underlying composition of these planetesimals. The purpose of this paper is to explain the observed compositions within the context of a model which includes the more realistic assumption that white dwarfs may be accreting multiple objects with a range of sizes. This new model is therefore based on that in \citet{2014MNRAS.439.3371W}.

In \citet{2017MNRAS.467.4970H}, the first large survey of polluted white dwarfs is presented with abundances found for multiple elements. Before this survey, there were only a handful of white dwarfs with multiple known metal abundances. This new survey therefore presents a unique opportunity to test prospective models statistically against a large sample. 

The observations are summarised in $\S$\ref{sec:Obs} and the model is described in $\S$\ref{sec:model}. The parameters that describe the composition of the accreted model are then constrained in $\S$\ref{sec:Results} under a variety of models. Our results are discussed in $\S$\ref{sec:Dis} which includes quantifying the accuracy of the assumption of single object accretion in $\S$\ref{sec:LO}, a discussion of the implications of the best fit model parameters in $\S$\ref{sec:CompFit} and predictions for time variation of the pollutant composition for white dwarfs with short sinking times in $\S$\ref{sec:Disc}. Finally, we present our conclusions in $\S$\ref{sec:Conc}.

\section{Observations}
\label{sec:Obs}

\subsection{The Earth as a Comparison}
\label{sec:SolSys}

Throughout the rest of this paper, comparisons will be made to the composition of the Earth and so the values used are introduced briefly here. The bulk Earth composition is formed of $f_{\oplus\text{,Ca}}=0.0171$, $f_{\oplus\text{,Mg}}=0.154$ and $f_{\oplus\text{,Fe}}=0.32$ by mass \citep{2003TrGeo...2..547M}. The composition of the Earth's crust, mantle and core are taken to be
\begin{equation}
    \begin{pmatrix}
    f_\text{Cru,Ca} & f_\text{Man,Ca} & f_\text{Cor,Ca} \\
    f_\text{Cru,Mg} & f_\text{Man,Mg} & f_\text{Cor,Mg} \\
    f_\text{Cru,Fe} & f_\text{Man,Fe} & f_\text{Cor,Fe}
    \end{pmatrix}
    =
    \begin{pmatrix}
    0.084 & 0.0253 & 0 \\
    0.062 & 0.228 & 0 \\
    0.065 & 0.0626 & 0.855
    \end{pmatrix}    
	\label{eq:Comp}
\end{equation}
which are obtained from \citet{2003TrGeo...2..547M} for the mantle and core and \citet{2014TrGeo...4..457W} for the (oceanic) crust and these compositions are shown in the ternary diagrams for the composition which follow.\footnote{The values used for the Earth's composition are slightly different to those used in \citet{2018MNRAS.477...93H}. Most notably, \citet{2018MNRAS.477...93H} use values for continental crust whereas we use oceanic crust.}

Denoting the fractions of the Earth's mass which is found in the crust, mantle and core as $f_{\oplus\text{,Cru}}$, $f_{\oplus\text{,Man}}$ and $f_{\oplus\text{,Cor}}$ respectively then the composition of each component and the bulk composition is linked via
\begin{equation}
	\begin{pmatrix}
    0.084 & 0.0253 & 0 \\
    0.062 & 0.228 & 0 \\
    0.065 & 0.0626 & 0.855
    \end{pmatrix}
    \begin{pmatrix}
    f_{\oplus\text{,Cru}} \\
    f_{\oplus\text{,Man}} \\
    f_{\oplus\text{,Cor}}
    \end{pmatrix}
    = \begin{pmatrix}
    0.0171 \\ 0.154 \\ 0.32
    \end{pmatrix}.
    \label{eq:bulkcomp}
\end{equation}
This can then be inverted to give
\begin{equation}
	\begin{pmatrix}
    f_{\oplus\text{,Cru}} \\
    f_{\oplus\text{,Man}} \\
    f_{\oplus\text{,Cor}}
    \end{pmatrix}
    = \begin{pmatrix}
    0.000148 \\ 0.675 \\ 0.325
    \end{pmatrix}.
    \label{eq:fractions}
\end{equation}

\subsection{White Dwarf Sample}
\label{sec:Samp}

\citet{2017MNRAS.467.4970H} present the first large sample of polluted white dwarfs where multiple elements are observed in each star, containing 231 DZ white dwarfs. These are white dwarfs which exhibit only metal lines (i.e. no hydrogen or helium lines) and so have He dominated atmospheres but temperatures below those required to observe He lines. The sample was selected using two methods. The first method primarily relies on a colour cut in $(u-g)$ and $(g-r)$ space which then required the removal of quasars and K/M dwarfs which are also found in the same region of colour space as the DZ white dwarfs. This was done by fitting to model spectra of K/M dwarfs and discarding observations that fit well K/M dwarfs. Finally, the remaining spectra were visually observed to identify the DZ white dwarfs.

The second (and more effective) method was to expand the sample from the first method by relaxing the colour cut used in the first method and fitting the observed spectra against both DZ model spectra and SDSS spectra of B--K main-sequence stars. Those that fit better to DZ model spectra were retained and those with a sufficiently good fit were then visually inspected to identify the DZ white dwarfs.

\begin{figure}
	\includegraphics[width=\columnwidth]{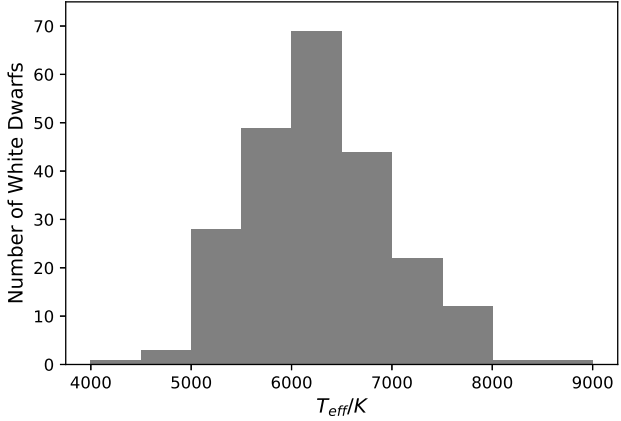}
    \caption{Histogram showing the effective temperature distribution of the 230 white dwarfs in the sample.}
    \label{fig:Sample}
\end{figure}

Of these 231 stars, one has a sufficiently large magnetic field that \citet{2018MNRAS.477...93H} were unable to produce a good fit to its spectrum, leaving a sample of 230 polluted white dwarfs. \citet{2017MNRAS.467.4970H} fit DZ model spectra to the observed spectra, allowing the determination of parameters such as the effective temperature and (importantly) the abundances (relative to the total mass of the convection zone) of the pollutant elements. Figure \ref{fig:Sample} shows the temperature distribution of these 230 white dwarfs. \citet{2017MNRAS.467.4970H, 2018MNRAS.477...93H} calculate the cooling ages of the white dwarfs as being between $1-8\,\text{Gyr}$ with the majority between $1-4\,\text{Gyr}$. Ca, Mg and Fe lines are observed in all 230 of these stars and so it is on these three elements that we will concentrate.

The errors in these abundances are estimated by \citet{2018MNRAS.477...93H} to be in the range of $0.05-0.3$ dex. We consider these errors in more detail in $\S$\ref{sec:Stel} and $\S$\ref{sec:Cav}.

For each of the 230 stars, sinking times are calculated for each pollutant element using models of the atmosphere dynamics and assuming that the dominant mechanism is gravitational settling of the metals out of the atmosphere. (For full details of the model and assumptions see \citealt{2017MNRAS.467.4970H}.) The distribution of sinking times across all 230 stars are $t_\text{sink,Ca}=10^{6.15\pm 0.13}\,\text{yr}$, $t_\text{sink,Mg}=10^{6.57\pm 0.09}\,\text{yr}$ and $t_\text{sink,Fe}=10^{6.12\pm 0.08}\,\text{yr}$. From this we can see that Mg is expected to sink between 2.5 and 3 times slower than Ca and Fe.

\begin{figure}
	\includegraphics[width=\columnwidth]{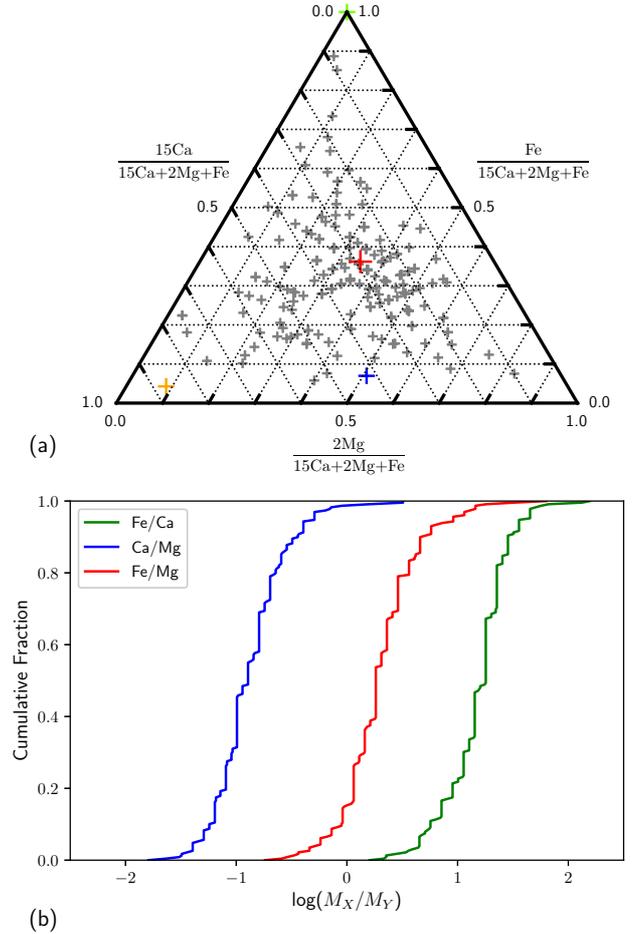}
    \caption{Compositional distribution measured in the white dwarf atmospheres. Panel (a) shows a ternary plot of the relative contribution by mass from Ca, Mg and Fe. Note the rescaling on Ca and Mg which has been done to help the display of the data. Also shown are bulk Earth (red), core Earth (green), mantle Earth (blue) and crust Earth (orange). Panel (b) displays the same data but in the form of three cumulative frequency plots for the three elemental mass ratios. Note that the ternary diagram in \citet{2018MNRAS.477...93H} is given in terms of number rather than mass. The reason that we use mass is that it is more natural when thinking about compositions of planetesimals.}
    \label{fig:Obs-Norm}
\end{figure}

With three elements, their relative abundances have two free parameters. Figure \ref{fig:Obs-Norm}a shows a ternary diagram of the relative mass abundances of Ca, Mg and Fe. The diagram shows that the sample appears to be centered around a bulk Earth composition, but with significant scatter (see also \citealt{2018MNRAS.477...93H}). Figure \ref{fig:Obs-Norm}b shows this same data but in the form of cumulative frequency plots. This form is useful as it allows the model outputs to be fit to the observed data.

\subsection{Standard Analysis}
\label{sec:StandAn}

While Figure \ref{fig:Obs-Norm} shows that the observed abundances in the stellar atmosphere appear to be centered about a bulk Earth composition, the different sinking times for the different elements (specifically the longer sinking time of Mg) mean that this does not necessarily imply that the accreted planetesimals will have a distribution that is also centered on bulk Earth.

When analysing the composition of white dwarf pollution, it is standard practice in the literature to assume that the pollution originates entirely from a single body (e.g. \citealt{2011ApJ...739..101Z,2013ApJ...766..132X,2015MNRAS.451.3237W,2018MNRAS.479.3814H}). This body is assumed to be processed through a disc which has a timescale of $t_\text{disc}$. This means that instead of the entire mass of the object arriving in the atmosphere instantaneously its arrival is instead spread out over $t_\text{disc}$. Since different elements sink on different timescales, the observed composition is not a constant over time. In this interpretation, there are broadly three distinct phases in the composition evolution and these are explored extensively in \citet{2009A&A...498..517K}. For two elements, $A$ and $B$, with respective sinking times $t_{\text{sink},A}$ and $t_{\text{sink},B}$ the three phases are:

\begin{enumerate}
\item For $t\ll t_\text{sink}$, differential sinking has not yet affected the composition and so the observed composition matches that of the accreted planetesimal
\begin{equation}
\left(\frac{A}{B}\right)_\text{Atm} = \left(\frac{A}{B}\right)_\text{Planetesimal}.
\label{eq:preSS}
\end{equation}
\item If $t_\text{disc}\gg t_\text{sink}$ then the accretion can settle into a steady state where the mass of pollutant in the atmosphere is approximately constant. This occurs after $\sim 5t_\text{sink}$ and leads to an observed composition of
\begin{equation}
\left(\frac{A}{B}\right)_\text{Atm} = \left(\frac{A}{B}\right)_\text{Planetesimal}\left(\frac{t_{\text{sink},A}}{t_{\text{sink},B}}\right),
\label{eq:steadystate}
\end{equation}
which can be simply inverted to give the planetesimal composition from the observed data \citep{2009A&A...498..517K}.
\item After the accretion has ceased the composition enters a declining phase. The composition then evolves with time according to
\begin{equation}
\left(\frac{A}{B}\right)_\text{Atm} = \left(\frac{A}{B}\right)_\text{Atm,0}\exp\left[t\left(\frac{t_{\text{sink},A}-t_{\text{sink},B}}{t_{\text{sink},A}t_{\text{sink},B}}\right)\right]
\end{equation}
where $t$ is the time from the end of accretion and $(A/B)_\text{Atm,0}$ is the atmospheric composition at $t=0$. The value of $(A/B)_\text{Atm,0}$ depends on the nature of the accretion event. If the composition reached steady state before declining then $(A/B)_\text{Atm,0}$ will equal the RHS of eq. \eqref{eq:steadystate}. If instead the accretion was virtually instantaneous (ie $t_\text{disc}\ll t_\text{sink}$ and each object enters the disc on a timescale that is less than the sinking timescale) then $(A/B)_\text{Atm,0}$ is simply the planetesimal composition as in eq. \eqref{eq:preSS}.
\end{enumerate}

\begin{figure}
	\includegraphics[width=\columnwidth]{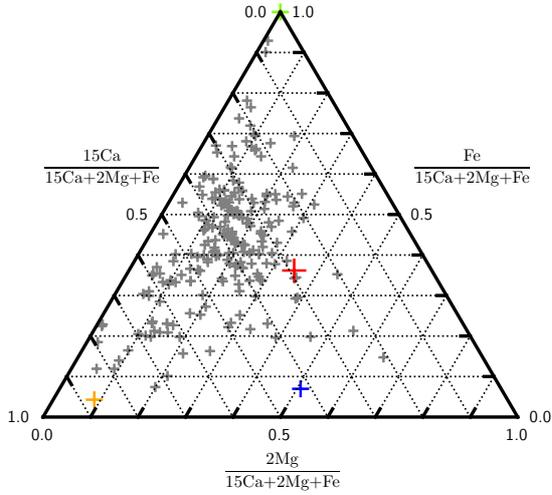}
    \caption{Compositional distribution as in Figure \ref{fig:Obs-Norm} panel (a) but now showing the required accreted composition, assuming that the accretion occurs in steady state.}
    \label{fig:Obs-SS}
\end{figure}

As well as showing the observed compositions, Figure \ref{fig:Obs-Norm} also shows the required planetesimal compositions if the accretion is observed in phase (i). Figure \ref{fig:Obs-SS} shows the required compositions under phase (ii) (steady state accretion) as in eq. \eqref{eq:steadystate}. Given that $t_\text{sink}$ is large for this sample, it is unlikely that any of these stars could be undergoing steady state accretion. Figure \ref{fig:Obs-Norm} should therefore be viewed as an indication of the effect of differential sinking on the observed composition. As expected, the required compositions have been shifted to lower values of Mg (i.e. upward and to the left), while the relative contributions of Fe and Ca are only negligibly affected. While there are still some observations with Mg enhanced relative to bulk Earth, the mean of the sample is no longer consistent with bulk Earth. Phase (iii) is time dependent and as such is too complicated to represent on these simple ternery diagrams.

\subsection{Stellar Composition}
\label{sec:Stel}

Some of the scatter in Figure \ref{fig:Obs-Norm} could originate in the underlying variation in the composition of stellar material. \citet{2016ApJS..225...32B} provide a sample of 1615 FGK stars for which the compositional data is known. Of these 1615 stars, 968 of them have a signal to noise ratio of greater than 100 and $\log(g)>3.5$. The other stars were discarded to avoid including non main-sequence stars and those with poor data (as in \citealt{2018MNRAS.479.3814H}).

\begin{figure}
	\includegraphics[width=\columnwidth]{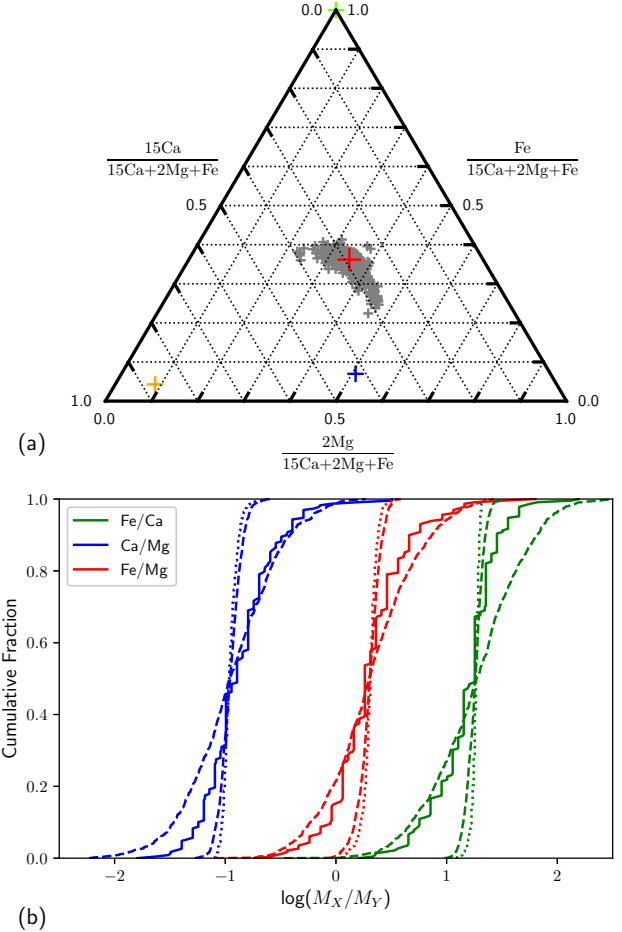}
    \caption{Panel (a) shows the composition of 968 FGK stars from \citet{2016ApJS..225...32B}. Panel (b) shows distributions for the observed data (solid), the stellar composition (dotted) and the stellar composition with random errors (dashed). The two dotted lines correspond to average errors of $0.05$ dex (narrower distribution) and $0.3$ dex (wider).}
    \label{fig:Obs-Stel}
\end{figure}

The composition of these 968 stars are shown in Figure \ref{fig:Obs-Stel}. The spread of points in Figure \ref{fig:Obs-Stel}a is clearly much less than in Figures \ref{fig:Obs-Norm} and \ref{fig:Obs-SS}. Figure \ref{fig:Obs-Stel}b shows distribution of the observed data (solid) and the stellar compositions (dotted), showing again that the stellar composition has a much narrower spread. However, as mentioned in $\S$\ref{sec:Samp}, the observed data has error estimates of between $0.05$ and $0.3$ dex. We have therefore added Gaussian random errors with $\sigma=0.05$ and $0.3$ dex to the stellar composition. The dashed lines in \ref{fig:Obs-Stel}b show the median distribution of these models.

We can see that errors of $0.05$ dex give a much smaller spread than the observations and so would not be able to explain the spread in the data. On the other hand, errors of $0.3$ dex are comparable to the spread in data. However, $0.3$ dex is the extreme value in the error range and so this is expected to be an overestimate of the true error. It is also worth noting that the errors only effect the spread of the distributions and do not change their mean. Without more specific details as to the errors on individual measurements it is difficult to implement then into our model. We therefore do not include the errors in our analysis but their effect on our conclusions is discussed in $\S$\ref{sec:Cav}.

\section{Description of the Model}
\label{sec:model}

Now that we have a large sample of polluted white dwarfs, we are in a position to be able to analyse the sample statistically. For this, we need a model that can track the atmospheric composition over time and from that produce a distribution for the predicted composition which can then be compared to the observations.

\subsection{Original Stochastic Accretion Model}

The model used throughout this work is based on that in \citet{2014MNRAS.439.3371W} and we briefly summarise that model here. This model was developed to explain the difference in observed accretion rates between white dwarfs with different sinking times. More specifically, white dwarfs with longer sinking times are inferred to have larger accretion rates (\citealt{2012ApJ...749..154G}; \citealt{2012MNRAS.424..464F}).

The model is a Monte Carlo simulation which tracks the mass remaining in the atmosphere of the white dwarf (and from which can be calculated an inferred accretion rate by dividing by the sinking time). The model assumes that planetesimals are being thrown into the white dwarf at a mass inflow rate $\dot{M}_\text{in}$ and that once the mass is in the atmosphere it decays exponentially on the sinking time $t_\text{sink}$. Therefore, a planetesimal of mass $m_\text{p,i}$ accreted at time $t_\text{p,i}$ will contribute
\begin{equation}
m_\text{atm,i} = m_\text{p,i}e^{-(t-t_\text{p,i})/t_\text{sink}}
\label{eq:decay}
\end{equation}
to the mass remaining in the atmosphere, where the subscript i refers to the mass from this single object. The total atmospheric mass is therefore given by
\begin{equation}
m_\text{atm} = \sum m_\text{atm,i}\, .
\label{eq:sum}
\end{equation}

If all the planetesimals are of the same mass, $m_p$ then the mean number of accretions per sinking time is
\begin{equation}
n = \dot{M}_\text{in}t_\text{sink}/m_p.
\label{eq:n}
\end{equation}
The simulation runs by using Poisson statistics to randomly choose the number of accreted bodies in one timestep and adding that to the mass in the previous timestep which is then decayed according to eq. \eqref{eq:decay}. The timestep is chosen to be $t_\text{sink}/10$ so that the shape of the exponential decay can be recovered.

A single value of $\dot{M}_\text{in}$ (in units of $\text{g}\,\text{s}^{-1}$) for all the simulations was not found to be compatible with the data so instead $\dot{M}_\text{in}$ was drawn from a log-normal distribution with parameters $\mu$ and $\sigma$. It was further found that the sinking times needed to be modified by including a disc timescale, $t_\text{disc}$. Since a full description of the disc is well beyond this model, the sinking timescale is instead replaced by a sampling timescale given by
\begin{equation}
t_\text{samp} = \sqrt{t_\text{disc}^2+t_\text{sink}^2}.
\label{eq:tsamp}
\end{equation}

Most notably for this work, it was found that mono-mass planetesimals (i.e. planetesimals all of the same mass) were incompatible with the observations. Therefore, a mass distribution was introduced which was parameterised as
\begin{equation}
n(m) \propto m^{-q},
\end{equation}
where $n(m)dm$ is the the number of objects in the range $m$ to $m+dm$. Under the assumption that $q<2$, the total mass is dominated by large objects and so the distribution is determined by $q$ and $m_\text{max}$, the mass of the largest object, only. This is implemented by splitting the mass distribution into $200$ logarithmically spaced mass bins down to an arbitrary small mass of $m_\text{min}=10^7\,\text{g}$, each of which has its own characteristic mass and mean accretion rate and is treated as a separate Poisson statistic with its own mean accretion rate. This rate is calculated as in eq. \eqref{eq:n} but with $m_p$ as the characteristic mass for each bin and $\dot{M}_\text{in}$ scaled for the proportion of the mass being input from each bin.

When using the model to try and fit to observations, the sinking times of each of the stars in the sample is used. Each star is run through the model $N_\text{com}$ times for $N_\text{tot}$ timesteps. Therefore, if there are $N_\text{star}$ stars in the sample then the total number of timesteps is $N_\text{star}N_\text{com}N_\text{tot}$. In each of the $N_\text{star}N_\text{com}$ simulations, $\dot{M}_\text{in}$ is randomly selected from its distribution.

The values concluded in \citet{2014MNRAS.439.3371W} are shown in Table \ref{tab:2014-values}. Unless explicitly stated otherwise, these values are used throughout the rest of this work.

At this point it is worth mentioning the tension between the disc value used here of $20\,\text{yr}$ and the values given in the majority of the literature (e.g. \citet{2012ApJ...749..154G} give $\log\left(t_\text{disc}/\text{yr}\right)=5.6\pm1.1$). This value is somewhat comparable with the sinking timescales of $\gtrsim10^6\,\text{yr}$. The potential consequences of this are discussed in $\S$\ref{sec:Cav}. Under the model developed in \citet{2014MNRAS.439.3371W}, disc timescales of $\sim 10^{5}\,\text{yr}$ were unable to provide good fits to the observations. Therefore, to ensure consistency within our model, we will continue with a disc timescale of $20\,\text{yr}$, as found in \citet{2014MNRAS.439.3371W}.

\begin{table}
    \centering
    \caption{Summary of the model values determined in \protect\citet{2014MNRAS.439.3371W}.}
    \label{tab:2014-values}
    \begin{tabular}{lccccc} 
        \hline
        Parameter & $\mu$ & $\sigma$ & $t_\text{disc}$ & $q$ & $m_\text{max}$ \\
        \hline
        Value & 8.0 & 1.3 & $20\,\text{yr}$ & 1.57 & $3.2\times 10^{24}\,\text{g}$ \\
        \hline
    \end{tabular}
\end{table}

\subsection{Inclusion of Composition}
\label{sec:Comp}

In this paper we extend this model by adding in the ability to track the composition of the accreted planetesimals. This requires a simple parameterisation of the possible compositions of the accreted planetesimals. To create this parameterisation, we will assume a relatively simple process by which the planetesimals are created. White dwarf pollutants can be analysed in terms of their mantle, core and crust fractions, assuming that they are similar to Earth (e.g. \citealt{2018MNRAS.479.3814H}), and so we will follow a similar method here. While this may not be an accurate description of the way planetesimals around white dwarfs are formed, it is nevertheless an illustrative way to consider the composition distribution.

Initially, we assume that a large parent body forms from a protoplanetary disc with metal composition (ie relative values of Ca, Mg and Fe) identical to that of its parent star. In reality, many parent bodies will form around the star, each with potentially different compositions. For simplicity of the model we will consider only one (which can be viewed as an average of all the parent bodies). The parent body then undergoes differentiation forming a core, mantle (and crust). There is evidence that destructive collisions are common in protoplanetary discs \citep{2016PEPS....3....7D} and that these continue well after any protoplanetary disc has vanished \citep{2008ARA&A..46..339W}. These collisions can lead to planetesimals whose bulk composition is different to the overall primitive composition of the parent body (\citealt{2015Icar..247..291B}; \citealt{2015ApJ...813...72C}). We therefore assume that the parent body undergoes collisional processing whereby the relative fractions of core, mantle and crust in the resultant planetesimals may differ from that of the parent body but the composition of the three components is unchanged.

For each simulation, one of the 968 FGK stars (from Figure \ref{fig:Obs-Stel}) is chosen to give the bulk composition (in Ca, Mg and Fe) of the parent body. In the first instance we shall ignore crust-like material and consider only core and mantle fractions. The overall composition of the parent body (PB) is related to the composition of the core and mantle via eq. \eqref{eq:compfraction}
\begin{equation}
	\begin{pmatrix}
    f_\text{Man,Ca} & f_\text{Cor,Ca} \\
    f_\text{Man,Mg} & f_\text{Cor,Mg} \\
    f_\text{Man,Fe} & f_\text{Cor,Fe}
    \end{pmatrix}
    \begin{pmatrix}
    f_\text{PB,Man} \\
    f_\text{PB,Cor}
    \end{pmatrix}
    = \begin{pmatrix}
    f_\text{PB,Ca} \\ f_\text{PB,Mg} \\ f_\text{PB,Fe}
    \end{pmatrix}
    \label{eq:compfraction}
\end{equation}
where $f_\text{Man,Ca}$ is the mass-fraction of the mantle which is Ca, $f_\text{PB,Man}$ is the fraction of the parent body which is mantle and $f_\text{PB,Ca}$ is the fraction of the parent body which is Ca.

Since the stellar compositions are given relative to H and we expect the parent body to form with much less H than the parent star, we renormalise these elemental compositions. To do this we assume (arbitrarily) that, $f_\text{PB,Fe}=f_{\oplus\text{,Fe}}=0.32$ and then use the stellar abundances relative to this to fix $f_\text{PB,Ca}$ and $f_\text{PB,Mg}$.

The body is then differentiated with a mantle fraction $f_\text{PB,Man}$ and a core fraction $f_\text{PB,Cor}=(1-f_\text{PB,Man})$. Finally, we assume that $f_\text{Cor,Fe}=0.855$ and $f_\text{Cor,Ca}=f_\text{Cor,Mg}=0$ as for the Earth's core (see eq. \ref{eq:Comp}). From this we then have sufficient information to find the values of $f_\text{Man,X}$.

Note that this process gives a lower limit to the value of $f_\text{PB,Man}$ as we require $0.855\times (1-f_\text{PB,Man})\leq 0.32$. In $\S$\ref{sec:Cru} crust-like material is added to the model. This is done as a subset of the mantle-like material and the precise way this is implemented is explained in $\S$\ref{sec:Cru} when it is needed.

This process only determines the composition and differentiation of the parent body. The model requires knowledge of the composition of the accreted planetesimals. As stated earlier, the planetesimals are assumed to form through collisional processing of the parent body. Further assumptions are required to determine how the differentiated fractions are combined in the planetesimals which ultimately gives their overall composition. This will be done in a few different ways in later sections and the methods will be introduced where needed.

\subsection{Independent Treatment of Elements in Atmosphere}

In addition to selecting the composition of each accreted planetesimal, once the mass is in the atmosphere, the different elements are treated independently and have the potential to sink out of the atmosphere at different rates (i.e. $t_\text{sink}$ could differ for different elements).

Due to the way the disc is incorporated into an overall sampling time, this model is only valid when $t_\text{sink}\gg t_\text{disc}$. In this case, the disc has only a negligible impact and so the modeling of the disc is unimportant. In the case that $t_\text{sink}\ll t_\text{disc}$ then it is the disc timescale which dominates. Therefore, eq. \eqref{eq:tsamp} shows that this implementation of the model results in all elements sinking at approximately the same rate. This means that the three distinct phases of evolution for a single object (see $\S$\ref{sec:StandAn}) would not be modelled correctly. The sample of 230 DZ stars considered here have sinking times of around a Myr and so the implementation of the disc (with its $20\,\text{yr}$ timescale) is unimportant. In $\S$\ref{sec:Disc} we present an adjustment to the model which correctly models the composition when $t_\text{sink}\lesssim t_\text{disc}$.

\begin{figure}
	\includegraphics[width=\columnwidth]{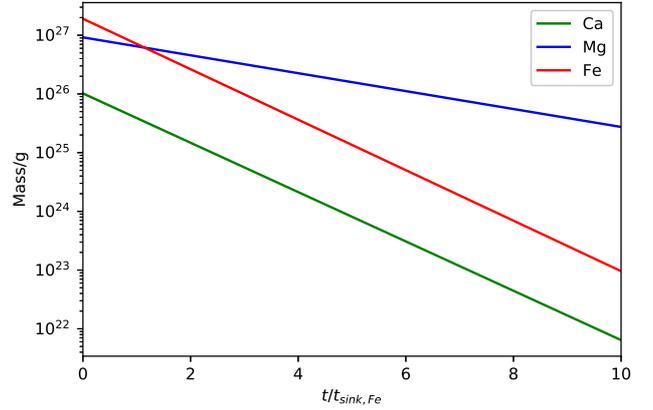}
    \caption{Results of using our model to track the pollutant mass of Ca, Mg and Fe after the accretion of a single Earth-like body onto the white dwarf SDSS$\,$J0002+3209. The expected exponential decays (with different timescales for each element) are seen.}
    \label{fig:Mod1}
\end{figure}

\begin{figure}
	\includegraphics[width=\columnwidth]{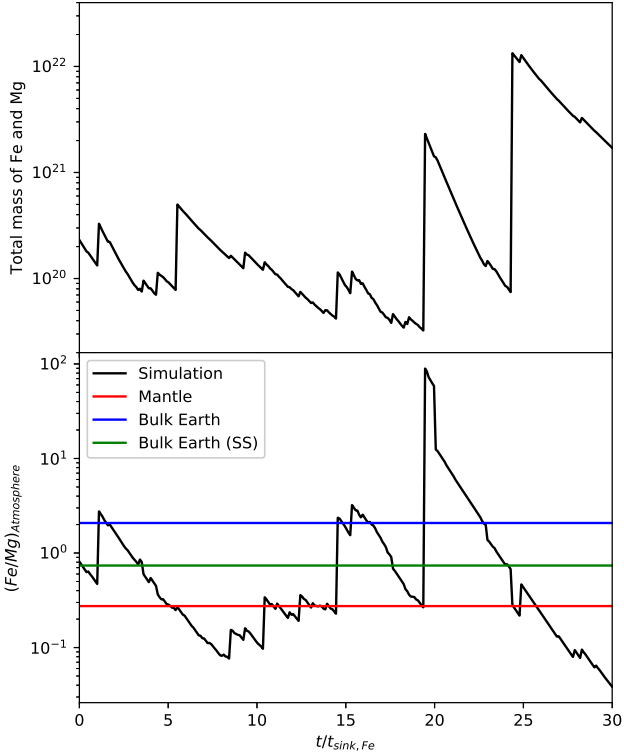}
    \caption{Results of using our model with a bimodal distribution. Each object is either mantle (with probability 0.675) or core. Top panel shows the total mass of Mg and Fe in the atmosphere while the bottom pane shows the Fe/Mg ratio in the atmosphere. Also shown are the mantle composition, the bulk Earth composition and the composition that would be observed in an object with a bulk Earth composition was being accreted in steady state. (Note that the core contains no Mg and so its Fe/Mg value is infinite.)}
    \label{fig:Mod2}
\end{figure}

Figure \ref{fig:Mod1} shows the model output for the accretion of a single Earth-like body with the sinking times appropriate for white dwarf SDSS$\,$J0002+3209 \citep{2017MNRAS.467.4970H}. These are $t_\text{sink,Ca}=10^{6.18}\,\text{yr}$, $t_\text{sink,Mg}=10^{6.62}\,\text{yr}$ and $t_\text{sink,Fe}=10^{6.17}\,\text{yr}$. These values are close to the average values given in $\S$\ref{sec:Samp} and so are representative of the sample as a whole. It shows that the accretion occurs essentially instantaneously before decaying exponentially. As expected, Ca and Fe decay at similar rates but Mg decays much more slowly. This will result in the observed composition becoming more Mg rich over time.

Figure \ref{fig:Mod2} shows the result of running the full model (i.e. with planetesimals accreted from a range of masses) for the same white dwarf as in Figure \ref{fig:Mod1}.  The composition for each object is taken randomly from a bimodal distribution where each accreted planetesimal is either Earth mantle-like (with probability 0.675) or Earth core-like (see eq. \eqref{eq:Comp}). The top panel shows the total mass of Mg and Fe (which is a proxy for the total mass) in which the accretion of a single massive object is evident as a large increase in mass following this event. Since $t_\text{sink}\gg t_\text{disc}$, there is no steady state phase and instead we immediately enter the declining phase where the mass decays exponentially.

The behaviour between accretion events is that of an exponential decline. However, as described in \citet{2014MNRAS.439.3371W}, there is also a quasi continuous level of accretion that comes from objects smaller than the crossover mass of
\begin{equation}
m_\text{tr} = m_\text{max}^{(2-q)/(1-q)}(\dot{M}_\text{in}t_\text{samp})^{1/(q-1)}.
\label{eq:mtr}
\end{equation}
The extent to which single objects dominate the observed pollution is explored in more detail in $\S$\ref{sec:LO}.

The bottom panel shows that each large accretion event is accompanied by a sudden change in the atmospheric composition towards that of the accreted object. In the case that the accreted object is mantle-like the composition changes to close to that of the Earth's mantle. If the accreted object is instead core-like then the composition will change to have an extremely high Fe/Mg value. This is because the accreted body has no Mg and so the Mg level in the atmosphere will be much lower than the Fe level. This is then followed by a declining phase where the composition changes exponentially (as expected) towards lower values of Fe/Mg.

This model is a natural improvement on the assumption that the observed pollution originates in the accretion of a single object used in the literature (see $\S$\ref{sec:StandAn}) since it reproduces that behaviour in the regime where that assumption is valid (see Figure \ref{fig:Mod1}), but also makes predictions about what would be observed in the more realistic case that multiple objects are contributing to the observed pollution.

\section{Results}
\label{sec:Results}

\begin{table*}
    \centering
    \caption{Summary of all models considered in $\S$\ref{sec:Results} along with their respective $\chi^2$ values.}
    \label{tab:Models}
    \begin{tabular}{cccccccccccc} 
        \hline
        Model & Multi/Sing & Comp Dist & Thresholds & Crust & Mg Dep & $t_\text{sink}$ & Fig & $\chi_\text{Fe/Ca}^2$ & $\chi_\text{Ca/Mg}^2$ & $\chi_\text{Fe/Mg}^2$ & $\sum\chi^2$ \\
        \hline
        A & Multiple & Bimodal & & & & Different & \ref{fig:Bin} & $140$ & $\infty$ & $1600$ & $\infty$ \\
        B & Multiple & Gaussian & & & & Different & \ref{fig:Gauss} & $12$ & $\infty$ & $1700$ & $\infty$ \\
        C & Multiple & Gaussian & \checkmark & & & Different & \ref{fig:Thresh} & $12$ & $\infty$ & $1400$ & $\infty$ \\
        D & Multiple & Gaussian & \checkmark & \checkmark & & Different & \ref{fig:Cru} & $13$ & $\infty$ & $1400$ & $\infty$ \\
        E & Multiple & Gaussian & \checkmark & \checkmark & \checkmark & Different & \ref{fig:MgDep} & $16$ & $25$ & $27$ & $68$ \\
        F & Multiple & Gaussian & \checkmark & \checkmark & \checkmark & Same & \ref{fig:Sink} & $25$ & $19$ & $27$ & $71$ \\
        G & Single (pSS) & Gaussian & N/A & \checkmark & \checkmark & N/A & \ref{fig:pSS} & $42$ & $16$ & $35$ & $92$ \\
        H & Single (SS) & Gaussian & N/A & \checkmark & \checkmark & Different & \ref{fig:SS} & $31$ & $19$ & $37$ & $86$ \\ \hline
    \end{tabular}
\end{table*}

We are now in a position to test both our model and the standard literature assumptions to try and fit their output to the observed data. To assess the quality of the fit to the data we will use the $\chi^2$ value for each of the cumulative frequency plots for the three elemental ratios (see Figure \ref{fig:Obs-Norm}b). For each ratio, the observed data are placed into logarithmically spaced bins of width 0.1. The model output is rescaled to give a total of 230 counts (equal to the number of observations) and divided into the same bins. To try and avoid problems of small number statistics, bins on the extreme ends of the distribution are combined to ensure that each bin has at least 5 observations. Strictly, this should be done with the expected rather than the observed counts. However, changing the bins for each combination of model parameters would alter the degrees of freedom and so not allow for direct comparison between parameters. Therefore, the combination is performed using only the observations and this gives 13 bins for each distribution (which are then fixed for all $\chi^2$ tests). The $\chi^2$ statistic is then calculated for each elemental ratio as
\begin{equation}
\chi^2 = \sum\frac{\left(N_\text{observed}-N_\text{model}\right)^2}{N_\text{model}},
\end{equation}
which can then be summed to give a total $\chi^2$ value.

Table \ref{tab:Models} shows the 8 different models used in the rest of this section, summarising the inclusions and assumptions of each one and the $\chi^2$ values of their best fit.

\subsection{Model A: Bimodal Composition}
\label{sec:Bim}

\begin{figure}
	\includegraphics[width=0.99\columnwidth]{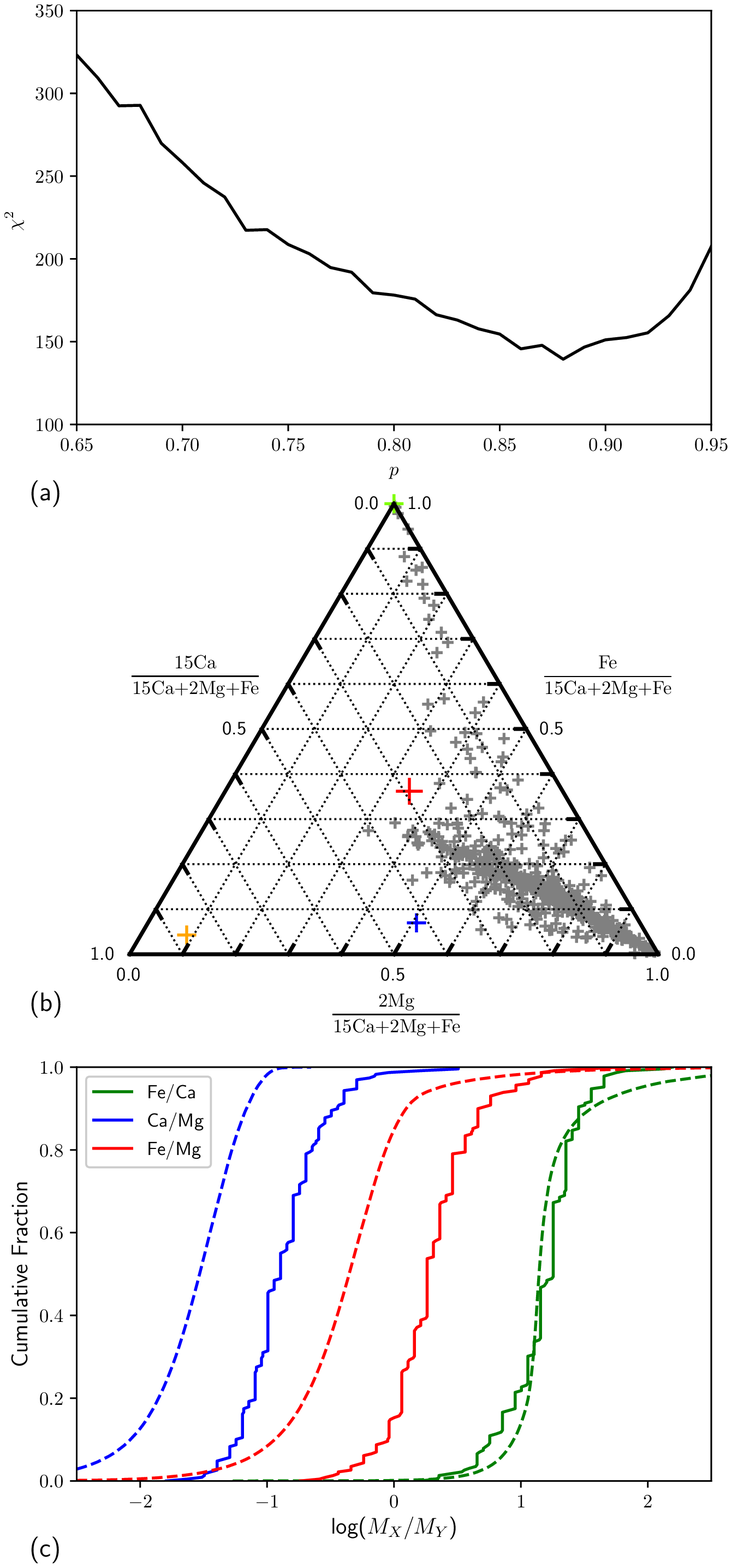}
    \caption{Model A - Results of the simulations with a bimodal composition distribution. Panel (a) shows the $\chi_\text{Fe/Ca}^2$ values for a range of $p$. Panel (b) shows 500 randomly selected points from the model output plotted on a ternary plot in the same way as in Figure \ref{fig:Obs-Norm}. Panel (c) shows cumulative frequency plots of the model output (dashed) and the observed data (solid) for the three elemental mass ratios. Panels (b) and (c) are created using the best fit of $p=0.88$.}
    \label{fig:Bin}
\end{figure}

In $\S$\ref{sec:Comp} we detailed how the compositions of the core and mantle components are calculated. To complete the model we require an assumption about the mantle fraction of the resultant planetesimals. The simplest assumption here is that each planetesimal is formed of either pure mantle-like material or pure core-like material. Since out of the total material a fraction $f_\text{PB,Man}$ is mantle-like, any randomly selected planetesimal must therefore be mantle-like with probability $p=f_\text{PB,Man}$ and core-like with probability $(1-p)$.

With such a model, for all values of $p$ the fit to Ca/Mg and Fe/Mg is so poor that it is only worth fitting to Fe/Ca. We therefore minimise the value of $\chi_\text{Fe/Ca}^2$ only. Figure \ref{fig:Bin}a shows the results of this for various values of $p$. The minimum value of $\chi_\text{Fe/Ca}^2$ is $140$ and corresponds to $p=0.88$. Figure \ref{fig:Bin}c shows that while the mean of the Fe/Ca appears to be well fitted by the model, the shape of the distribution does not match that of the observations. Further to this, the Ca/Mg and Fe/Mg plots show that the model distributions are much lower than the observed values. This corresponds to a lack of Mg in the observations as predicted earlier.

Figure \ref{fig:Bin}b shows the results of the simulation on the ternary plot. The majority of the points lie around a line from the bottom right corner (pure Mg) to a point just below the bulk Earth composition. Lines of constant Fe/Ca are straight lines running from the bottom right corner. In this case, the mantle composition used in the model lies just below the bulk Earth composition at the left-hand end of this line. Since $p=0.88$ is large, the core fraction is very small and so more Fe ends up in the mantle than in Earth's mantle.

The points lying on and around this line are dominated by a single mantle-like object that has been in the atmosphere for several sinking times. Objects that have been in the atmosphere for longer are found closer to the bottom right corner as they have undergone a longer period of differential sinking. The small variation around this line is caused by three factors. The small difference in the sinking times of Fe and Ca, the small difference in the mantle composition due to the variation in the composition of the parent star (see Figure \ref{fig:Obs-Stel}) and the amount of core-like material that has been accreted alongside the mantle-like material. We can also see some core-like compositions near the top corner as well as some that lie in between, corresponding to various mixtures of core and mantle-like compositions. It is clear that this plot is nothing like the data in Figure \ref{fig:Obs-Norm} and also can never be by construction. This is because our model provides no mechanism by which objects can move to the left (lower Mg) than the central line connecting the mantle and core compositions. Note that, due to the slight variation in stellar composition used in the simulations (as shown in Figure \ref{fig:Obs-Stel}), each simulation has a slightly different mantle composition and so observations can lie slightly to the left of the line connecting the Earth mantle and core compositions.

\subsection{Model B: Logit-Normal Composition}
\label{sec:Gauss}

\begin{figure}
	\includegraphics[width=\columnwidth]{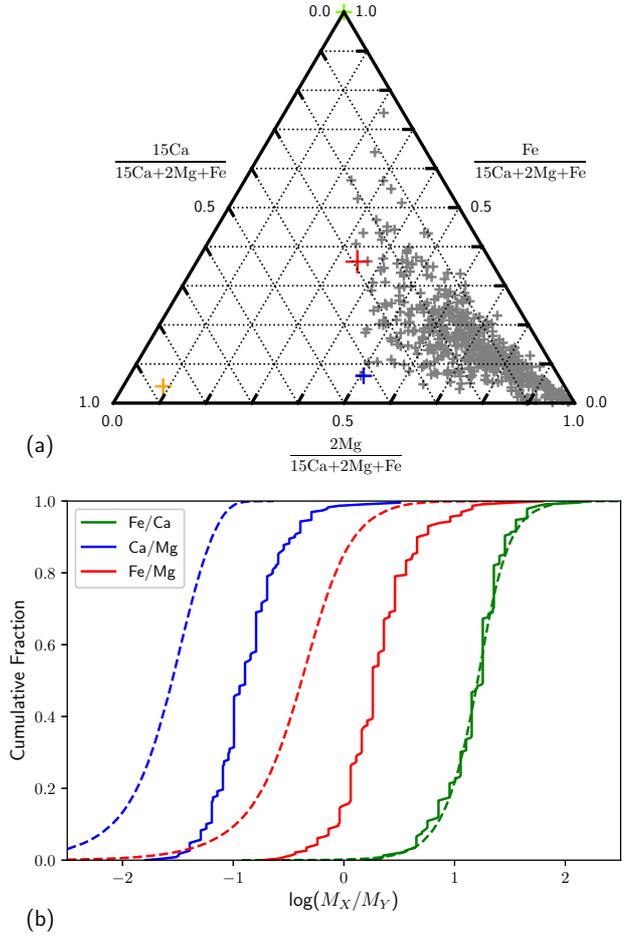}
    \caption{Model B - Results of the simulations with a logit-normal composition distribution, created using the best fit of $\mu_\text{man}=0.27$ and $\sigma_\text{man}=0.42$.}
    \label{fig:Gauss}
\end{figure}

With the bimodal parameterisation (with one free parameter) being unable to fit to any of the three distributions, the next simplest model would have two free parameters. In this model, each planetesimal is a mix of both mantle and core-like material. To do this, we choose a logit-normal distribution (base $10$) for the mantle fraction of each planetesimal, $f_\text{plan,Man}$, such that $\log\left(\frac{f_\text{Plan,Man}}{1-f_\text{Plan,Man}}\right)$ follows a normal distribution when mean and standard deviation $\mu_\text{Man}$ and $\sigma_\text{Man}$. Note that the value of $\mu_\text{Man}$ is not the average mantle fraction of the planetesimals. A value of $\mu_\text{Man}=0$ corresponds to $\left\langle f_\text{plan,Man}\right\rangle=0.5$ and negative values of $\mu_\text{Man}$ are allowed and give $\left\langle f_\text{plan,Man}\right\rangle<0.5$.

The reason for this choice of distribution is that it ensures that $0<f_\text{plan,Man}<1$ which is required for the mantle fraction of an object. To ensure that the bulk composition is equal to that of the parent star, $f_\text{PB,Man}$ can be found numerically as the mean value of $f_\text{Plan,Man}$ and can then be used to determine the composition of the core and mantle.

For computational efficiency, we make a simplifying approximation for any of the $200$ mass bins that accretes $>10$ objects in any given timestep. In this case, the total accreted composition is assumed to be the mean composition of the distribution. While this seems like a major simplification, this still leaves $\sim90$ objects in each timestep which are treated properly. Since the objects are predominantly those that are arriving in mass bins with low accretion rates, they will be the large objects which dominate the total mass budget. The value of $10$ for the approximation was varied to test its effect on the model output. No significant change was seen if it was increased, confirming that this approximation does not affect our results while saving significantly on computational expense.

Once again, the fits to Ca/Mg and Fe/Mg are so poor that we only consider Fe/Ca. With this, we repeat the $\chi_\text{Fe/Ca}^2$ procedure as before, now minimising over the 2D space of $\mu_\text{Man}$ and $\sigma_\text{Man}$.

The minimum value of $\chi_\text{Fe/Ca}^2$ is now found to be $12$ which is given by $\mu_\text{Man}=0.27$, $\sigma_\text{Man}=0.42$ which corresponds to $f_\text{PB,Man}=0.63$. The results of this fit are shown in Figure \ref{fig:Gauss}. From Figure \ref{fig:Gauss}b we can see that there is now a good fit to the Fe/Ca line. As before, the model values of Ca/Mg and Fe/Mg are lower than the observed values.

Figure \ref{fig:Gauss}a shows the ternary plot of the model output. Comparing this with that in Figure \ref{fig:Bin} we see that the model compositions are now more widely distributed, indicative of the greater variation in the accreted composition. The same trend is seen with observations being found from near the accreted composition (near the line connecting the core and mantle compositions) to an observed composition that is almost pure Mg.

\subsection{Model C: Inclusion of Detection Thresholds}
\label{sec:thresh}

\begin{figure}
	\includegraphics[width=\columnwidth]{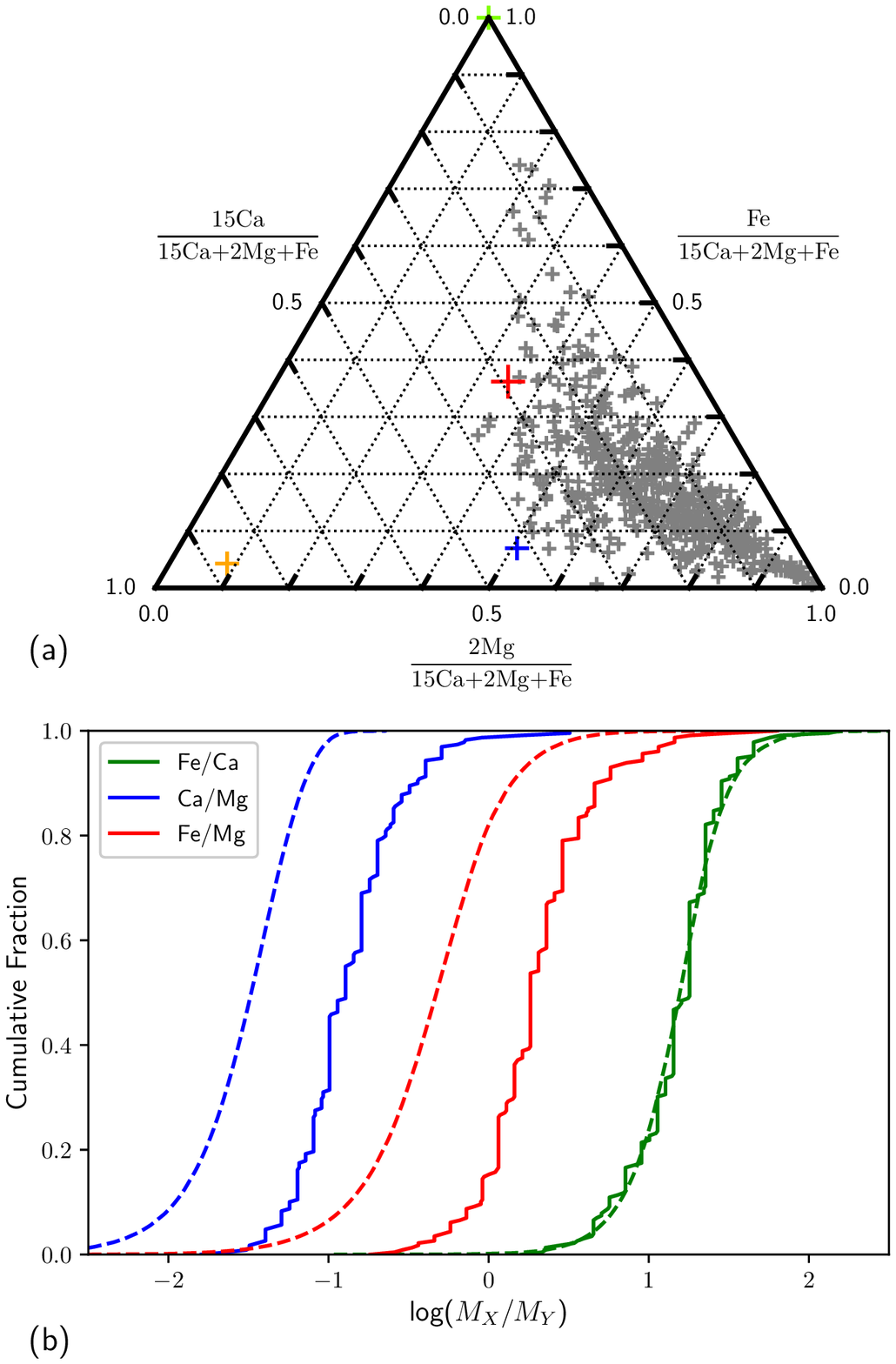}
    \caption{Model C - Results of the simulations with a Gaussian composition distribution and the thresholds given in Table \ref{tab:thresh}. Apart from the thresholds, the plots are created with the same parameters as Figure \ref{fig:Gauss}.}
    \label{fig:Thresh}
\end{figure}

Thus far, the analysis of the model output has included all of the timesteps regardless of the levels of each of the elements in the atmosphere. However, the sample only included DZ white dwarfs (i.e. those with observed metal lines). White dwarfs in the same temperature range as this DZ sample (see Figure \ref{fig:Sample}) but with very low levels of pollutant would not have observable metal lines and so would be excluded. Since the pollutant elements have different sinking times it is plausible that stars with low pollutant levels could have different observed compositions to those with higher levels.

A full and proper treatment of this would involve calculating the pollutant level required for each white dwarf to have observed metal lines (and so be included in the sample) individually. Then, for each star run through the simulation, any values that fall below the threshold would be removed from the model output. As the detection thresholds are likely to vary between stars (based on their temperature and magnitude), calculating them for all 230 stars is not a simple procedure. Instead, here we estimate a simple detection threshold to test its significance.

To find an approximate detection threshold (in terms of the mass in the atmosphere) we first estimate the mass in the convection zone of each element for each of the 230 stars. \citet{2017MNRAS.467.4970H} give the size of the convection zone as a fraction of the white dwarf's total mass and the abundances of each element. If we assume a typical white dwarf mass of $0.6\,\text{M}_{\sun}$ then we can find an estimate for the mass of each element in the atmosphere. We then use the smallest of these masses as the detection threshold. This gives a reasonable order of magnitude estimate and the results of these thresholds are shown in Table \ref{tab:thresh}.

\begin{table}
    \centering
    \caption{Estimates of the detection thresholds for Ca, Mg and Fe for the sample of 230 white dwarfs.}
    \label{tab:thresh}
    \begin{tabular}{lccc} 
        \hline
        Element & Ca & Mg & Fe \\
        \hline
        Threshold/g & $2.9\times 10^{18}$ & $1.3\times 10^{19}$ & $2.3\times 10^{19}$ \\
        \hline
    \end{tabular}
\end{table}

Adding these thresholds to the logit-normal model results in discarding approximately half of the model output in any one simulation. The result of running the simulation with these thresholds is shown in Figure \ref{fig:Thresh}. In comparison with Figure \ref{fig:Gauss} we can see that the threshold does not remove all compositions with equal likelihood. With the threshold, the model predictions for Ca/Mg and Fe/Mg have been shifted to slightly larger values. This can also be seen in Figure \ref{fig:Thresh}a where there are fewer points in the bottom right hand corner than in Figure \ref{fig:Gauss}a.

To understand why these points with extremely high relative Mg abundances are preferentially removed, we need to consider the situations in which these extreme abundances occur. Considering a single object, the longer it has spent in the atmosphere the more enhanced the Mg will be relative to Ca and Fe. Therefore, for the observed composition to have an extreme Mg abundance, it must be dominated by an object which was accreted many sinking times ago. However, these objects will also have been depleted to a level where the total mass remaining in the atmosphere will have decreased by several orders of magnitude and will therefore be much less likely to exceed the threshold than when it had just been accreted. Therefore, the compositions with high relative Mg abundances will be preferentially removed compared to the other compositions.

Further to this, the Fe threshold is higher than the Mg one. Therefore, even for compositions for which Fe/Mg is near unity if the mass of each element is around $1.5\times 10^{19}\,\text{g}$ then the thresholds will select those with larger values of Fe/Mg.
This will further increase the average value of Fe/Mg as seen.

The implementation of a crude observational threshold into the model has reduced the disagreement between the model output and the observed data. We should therefore consider whether small adjustments to the thresholds in Table \ref{tab:thresh} can solve the disagreement completely. However, altering the values by two orders of magnitude has no major effect on the plots in Figure \ref{fig:Thresh}. If the limits are changed more than this then more than $99\%$ of the model output are removed and the model output starts to suffer from small number statistics. It therefore does not appear as if adjustment of the thresholds alone can reconcile the model output and the observed data. However, as the thresholds do affect the model outputs they will be left in the model and used in all future simulations.

\subsection{Model D: Inclusion of Crust-like Material}
\label{sec:Cru}

\begin{figure}
	\includegraphics[width=\columnwidth]{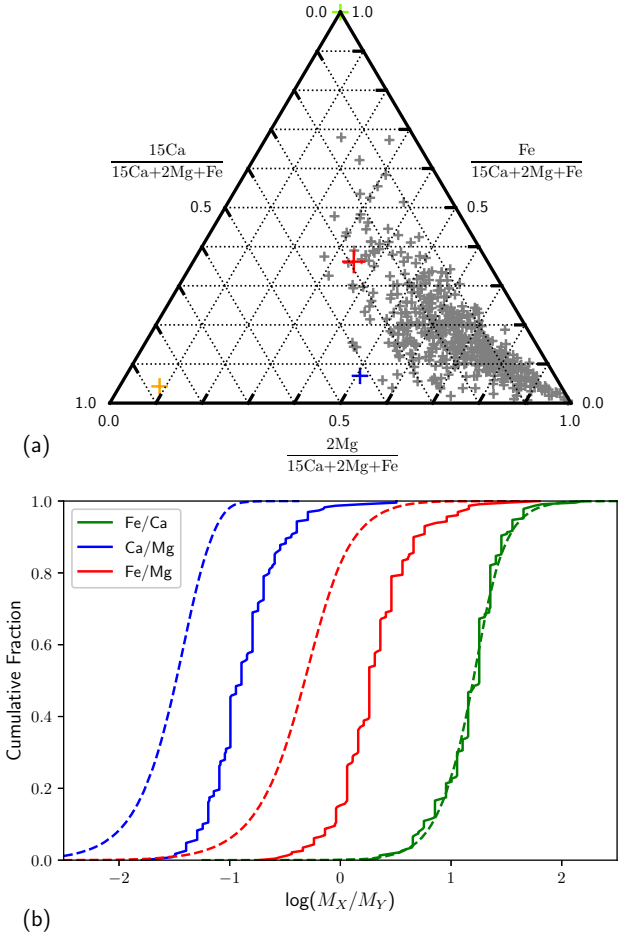}
    \caption{Model D - Results of the simulations with a logit-normal composition distribution including the thresholds given in Table \ref{tab:thresh} and the introduction of a crust-like fraction. The parameters used are the best fit paramemeters from Figure \ref{fig:Gauss} with the addition of $\mu_\text{Cru}=-1.5$ and $\sigma_\text{Cru}=0.5$.}
    \label{fig:Cru}
\end{figure}

Thus far we have not included any crust-like material and so it is now worth asking whether its inclusion can improve the fit of the Ca/Mg and Fe/Mg lines. Indeed, the presence of crust-like material would provide a mechanism whereby the observed pollutant compositions could be found in the left half of the ternary diagrams. We therefore allow for each accreted body to be formed of not just core and mantle-like material but also crust-like material.

The crustal material is parameterised as a subset of the mantle. For each body, the proportion of mantle that is instead crust is chosen from a logit-normal parameterised by $\mu_\text{Cru}$ and $\sigma_\text{Cru}$. The reason that we chose crust as a subset of mantle is because this means that bodies with higher mantle fractions are likely to have higher crust fractions. This makes good physical sense as collisional models do not reproduce bodies with high core and crust fractions but very little mantle (e.g. \citealt{2018E&PSL.484..276C}). Under this parameterisation these scenarios are very unlikely. It is worth noting that $\mu_\text{Man}$ and $\sigma_\text{Man}$ now parameterise the total fraction of mantle and crust.

With both crust and mantle material, eq. \eqref{eq:compfraction} becomes
\begin{equation}
	\begin{pmatrix}
    f_\text{Cru,Ca} & f_\text{Man,Ca} & f_\text{Cor,Ca} \\
    f_\text{Cru,Mg} & f_\text{Man,Mg} & f_\text{Cor,Mg} \\
    f_\text{Cru,Fe} & f_\text{Man,Fe} & f_\text{Cor,Fe}
    \end{pmatrix}
    \begin{pmatrix}
    f_\text{PB,Cru} \\
    f_\text{PB,Man} \\
    f_\text{PB,Cor}
    \end{pmatrix}
    = \begin{pmatrix}
    f_\text{PB,Ca} \\ f_\text{PB,Mg} \\ f_\text{PB,Fe}
    \end{pmatrix}.
    \label{eq:compfraction2}
\end{equation}
As before we take $f_\text{PB,Fe}=f_{\oplus\text{,Fe}}=0.32$ and then use the stellar abundances relative to this to fix $f_\text{PB,Ca}$ and $f_\text{PB,Mg}$. We can find $f_\text{PB,Cor}$, $f_\text{PB,Man}$ and $f_\text{PB,Cru}$ from the means of the logit-normal distributions. We again take $f_\text{Cor,Fe}=0.855$ and $f_\text{Cor,Ca}=f_\text{Cor,Mg}=0$ as in $\S$\ref{sec:Comp}.

We now need the values of $f_\text{Man,X}$ and $f_\text{Cru,X}$. To do this we assume that the ratios of each element in the mantle and crust are equal to those for the Earth
\begin{equation}
	\frac{f_\text{Man,X}}{f_\text{Cru,X}} = \left(\frac{f_\text{Man,X}}{f_\text{Cru,X}}\right)_\oplus
    \label{eq:ManCruRatio}
\end{equation}
which then gives sufficient information to find all values of eq. \eqref{eq:compfraction2}.

There are now four free parameters in our model. However, varying these parameters cannot provide a good fit to the Ca/Mg or Fe/Mg lines. The reason for this is that, while we have introduced crust-like material into our model, the average accreted  composition is still set by the composition of the FGK stars. Figure \ref{fig:Cru} shows the results of running the new model with the best fit parameters used in Figure \ref{fig:Gauss} with the addition of $\mu_\text{Cru}=-1.5$ and $\sigma_\text{Cru}=0.5$. These parameters correspond to $f_\text{Cru}=0.03$, $f_\text{Man}=0.59$ and $f_\text{Cor}=0.37$.

Figure \ref{fig:Cru} shows a few points now lying slightly towards a crust-like composition but that the overall change has been minimal. This is not surprising given that the total crust fraction is $<5\%$ and its addition has not changed the average accreted composition.

\subsection{Model E: Inclusion of Magnesium Depletion}
\label{sec:MgDep}

\begin{figure}
	\includegraphics[width=\columnwidth]{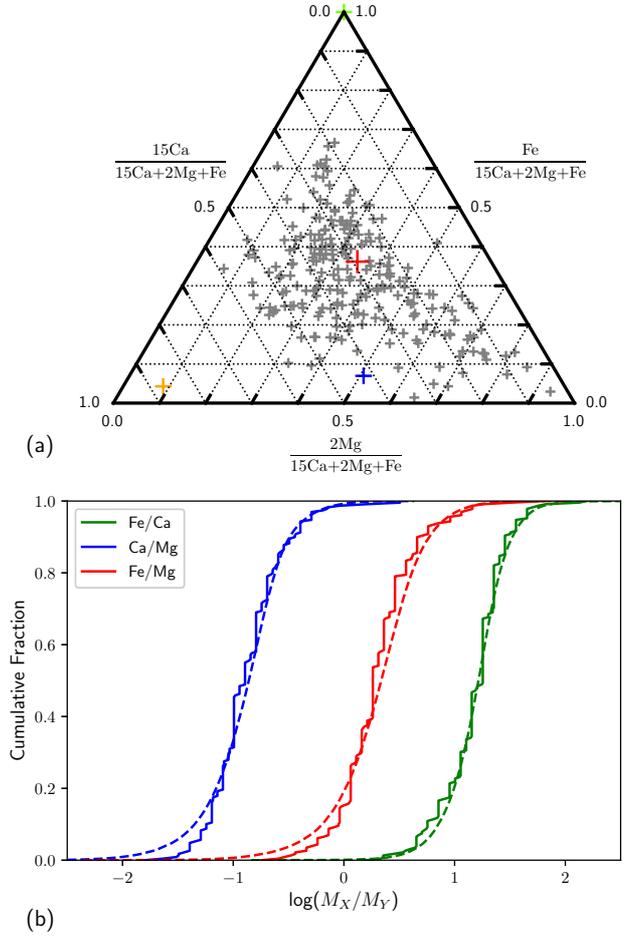}
    \caption{Model E - Results of the simulations with a logit-normal composition distribution including the thresholds given in Table \ref{tab:thresh}, the inclusion of a crust-like fraction and a systematic Mg depletion. The parameters used are those given in Table \ref{tab:Parameters}.}
    \label{fig:MgDep}
\end{figure}

The major discrepancy we have lies in the apparent lack of Mg in the data compared to our model output. This can be shown by considering the ratio of the mean abundances of our model compared to the data. These values are $1.05$ for Fe/Ca, $0.19$ for Ca/Mg and $0.23$ for Fe/Mg.

Since it does not appear that observational biases can completely explain the discrepancy we must consider the possibility that it is a real, physical effect. In this section we will introduce a parameterisation for a Mg depletion in the planetesimal material. We do this initially without considering a physical explanation but this will be considered in $\S$\ref{sec:Dis}.

The simplest parameterisation is to assume that the Mg is given by the stellar Mg multiplied by a depletion factor $f_\text{Dep,Mg}$ (i.e. $f_\text{PB,Mg}$ is multipied by $f_\text{Dep,Mg}$). Physically, $f_\text{Dep,Mg}$ cannot be negative and could in some cases be greater than $1$ (in which case Mg is enhanced). We allow $f_\text{Dep,Mg}$ to be different in each of our $N_\text{star}N_\text{com}$ simulations and draw its value randomly from a log-normal distribution parameterised by $\mu_\text{Mg}$ and $\sigma_\text{Mg}$. This choice of distribution means that $f_\text{Dep,Mg}$ can range from $0$ to $\infty$. (Although a constant depletion by a fixed value has only one parameter, this is covered by the case $\sigma_\text{Mg}=0$.)

We now have 6 free parameters for the mantle fraction, crust fraction and Mg depletion. We therefore now need to fit over all 6 parameters. To do this, we calculate a binned $\chi^2$ for each of the three distributions and minimise the sum of these. The minimum total value of $\chi^2$ was found to be 68 (with individual values of $\chi^2$ for the three distributions given in Table \ref{tab:Models}) and is given by the specific values of the parameters in Table \ref{tab:Parameters}.

The results of this model are shown in Figure \ref{fig:MgDep}. We can see that the inclusion of the Mg depletion has drastically improved the agreement between the model and the data. One particular point of note lies in the tails at the low end of the Ca/Mg and Fe/Mg distributions. These tails correspond to pollution with a relatively high Mg abundance. A large amount of this disagreement is due to the differential sinking times. The longer sinking time for Mg creates the long tails of Mg enriched observations. These can be seen from Figure \ref{fig:Gauss}b onwards. The thresholds introduced in $\S$\ref{sec:thresh} reduced the tails slightly but insufficiently to provide a good fit between the model output and the data. Despite this disagreement, the model is able to produce much of the behaviour seen in the observations.

The value of $f_\text{PB,Cru}=0.15$ which results from our best fit parameters is very large for a crust fraction. The reason for such a large value is that the large values of Ca/Mg in the observations require bodies with large crust fractions to reproduce them. One way to address this would be to adjust our model. Intuitively, we might expect that pure crust-like objects would be a possible result of the collisional processes which form the planetesimals and several are reported in the analysis of \citet{2018MNRAS.477...93H} on this data. Under our model, this scenario requires first a combined mantle and crust fraction of close to $1$ and also a crust fraction (of the crust and mantle combination) of close to $1$ which is relatively unlikely. Instead, we could imagine adding a parameterisation for the crust fraction whereby objects which have large mantle fractions have larger values of $\mu_\text{Cru}$ (and potentially $\sigma_\text{Cru}$) than those with small mantle fractions. However, even a simple linear parameterisation of $\mu_\text{Cru}$ and $\sigma_\text{Cru}$ would increase the number of parameters in our model to 8. Due to computational limits and a desire to keep our model as simple as possible, we do not include these parameters in the model.

\begin{table*}
    \centering
    \caption{Values of the 6 parameters that give the best fit to all of the data under Models E to H along with the total $\chi^2$ for the models.}
    \label{tab:Parameters}
    \begin{tabular}{lccccccccccc} 
        \hline
        Model & $\mu_\text{Man}$ & $\sigma_\text{Man}$ & $\mu_\text{Cru}$ & $\sigma_\text{Cru}$ & $\mu_\text{Mg}$ & $\sigma_\text{Mg}$ & $\chi^2$ & $f_\text{PB,Cru}$ & $f_\text{PB,Man}$ & $f_\text{PB,Cor}$ & $\left\langle f_\text{Dep,Mg}\right\rangle$ \\
        \hline
        E & $0.29$ & $0.36$ & $-0.90$ & $0.95$ & $-0.65$ & $0.00$ & $68$ & $0.15$ & $0.49$ & $0.36$ & $0.22$ \\
        F & $0.32$ & $0.50$ & $-0.90$ & $0.85$ & $-0.06$ & $0.25$ & $71$ & $0.15$ & $0.49$ & $0.36$ & $1.03$ \\
        G & $0.25$ & $0.28$ & $-1.2$ & $0.74$ & $-0.07$ & $0.23$ & $92$ & $0.08$ & $0.55$ & $0.37$ & $0.98$ \\
        H & $0.26$ & $0.27$ & $-1.0$ & $0.52$ & $-0.50$ & $0.28$ & $86$ & $0.09$ & $0.55$ & $0.37$ & $0.39$ \\
        \hline
    \end{tabular}
\end{table*}

\subsection{Model F: Adjustment to Sinking Times}
\label{sec:Sink}

As noted earlier, the reason for the systematic offset seen between the model and observations in the Ca/Mg and Fe/Mg distributions is the longer sinking time for Mg compared to Ca and Fe. In $\S$\ref{sec:MgDep} this was addressed by supposing that the accreted planetesimals are systematically depleted in Mg compared to their parent star.

In this section we propose another possible explanation that the sinking times given in \citet{2017MNRAS.467.4970H} are incorrect. One potential reason for suspecting this to be the case is to look at the sinking times given in \citet{2009A&A...498..517K}. Although these are older calculations, they show that Ca, Mg and Fe sink on much more similar timescales (with some temperature dependence). For example, for a $6000\,$K DB white dwarf (a typical star for our sample) the sinking times are $t_\text{sink,Ca}=10^{6.56}\,\text{yr}$, $t_\text{sink,Mg}=10^{6.55}\,\text{yr}$ and $t_\text{sink,Fe}=10^{6.56}\,\text{yr}$.

As we are not in a position to calculate new sinking times for all 230 of the stars in our sample the model is instead adjusted in a very simple way. Instead of each element sinking on its own timescale the Mg and Fe sinking timescales are replaced with that for Ca so that all elements sink on the same timescale. The same fit as in $\S$\ref{sec:MgDep} is then performed. Note that a possible Mg depletion is left in to see if it is necessary to produce a good fit to all three distributions.

\begin{figure}
	\includegraphics[width=\columnwidth]{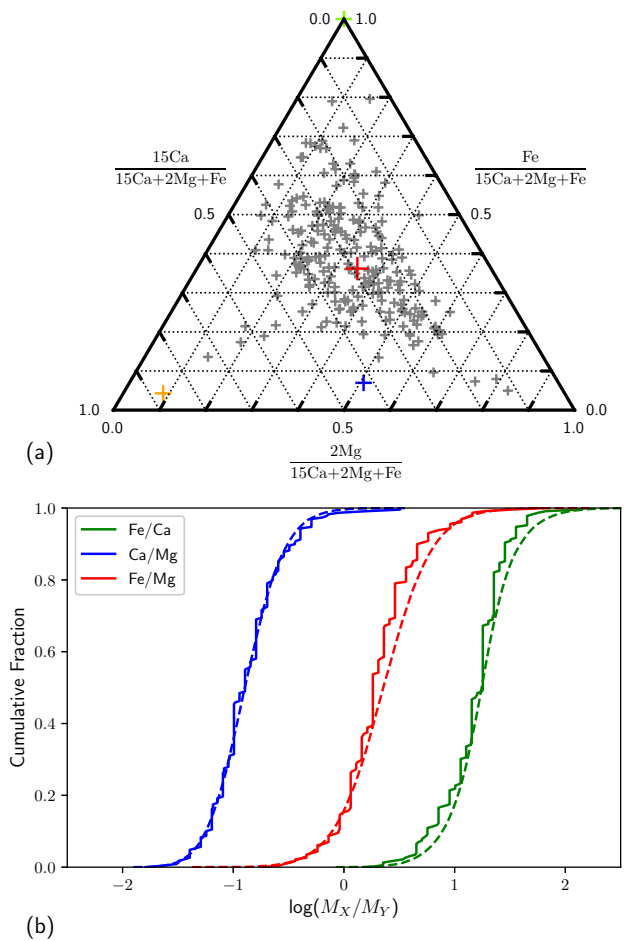}
    \caption{Model F - Results of the simulations with a Gaussian composition distribution including the thresholds given in Table \ref{tab:thresh}, a systematic Mg depletion, a crust-like fraction but with all elements sinking on the same timescale. The parameters used are those given in Table \ref{tab:Parameters}.}
    \label{fig:Sink}
\end{figure}

The minimum total value of $\chi^2$ is now found to be $71$ (with the specific values given in Table \ref{tab:Models}) and is given by the parameters in Table \ref{tab:fractions}. The value of $\chi^2$ is very similar value to that in $\S$\ref{sec:MgDep}. The results for this best fit are given in Figure \ref{fig:Sink}. In comparison to Figure \ref{fig:MgDep} we can see that, as we might expect, the discrepancy at low values of Ca/Mg and Fe/Mg has now disappeared. This is because there is no longer any differential sinking which can cause these extreme abundance ratios. Similarly, looking at the ternary plots we see that Figure \ref{fig:Sink} has fewer points in the lower right corner of the plot. These are the Mg enriched points which we would expect to be absent when the sinking times are the same.

Comparing the best fit parameters seen in Table \ref{tab:Parameters} we see that the parameters for the mantle and crust fractions are similar between the two models. As expected, the Mg depletion parameters differ drastically between the two. The value of $\mu_\text{Mg}$ has increased from $-0.65$ in Model E to $-0.06$ in Model F. This is because Model F does not have differential sinking times and so Mg is not enhanced in the atmosphere over time. It is also notable that $\sigma_\text{Mg}=0.25$. The reason for this is that before, the differential sinking times were creating a spread in the model output for Mg (compared to Ca and Fe). With identical sinking times, this spread in the data must be created with a large value of $\sigma_\text{Mg}$.

\subsection{Models G and H: Standard Analysis}

\begin{figure}
	\includegraphics[width=\columnwidth]{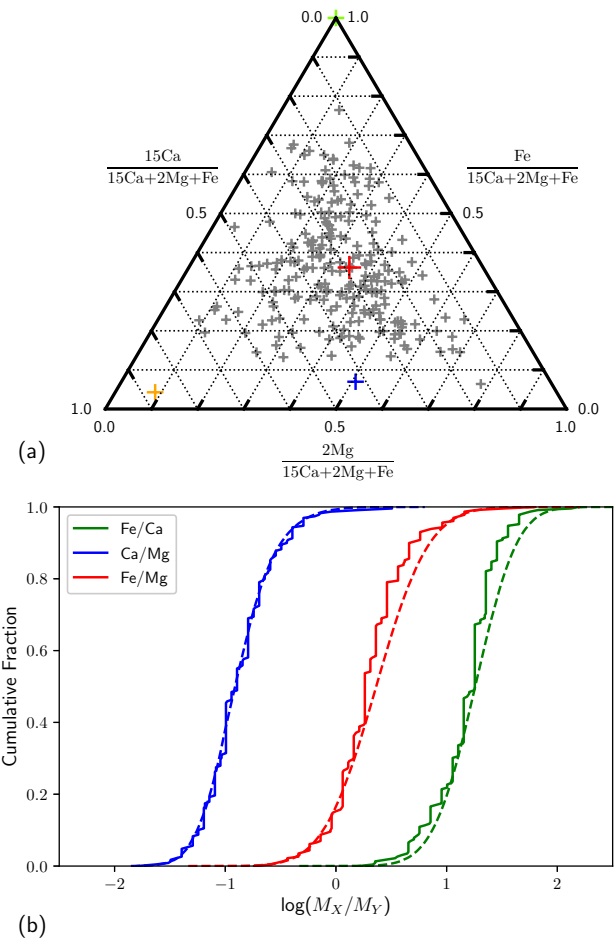}
    \caption{Model G - Best fit to the observations assuming that all the model compositions are given by single objects in the pre-steady state phase.}
    \label{fig:pSS}
\end{figure}

\begin{figure}
	\includegraphics[width=\columnwidth]{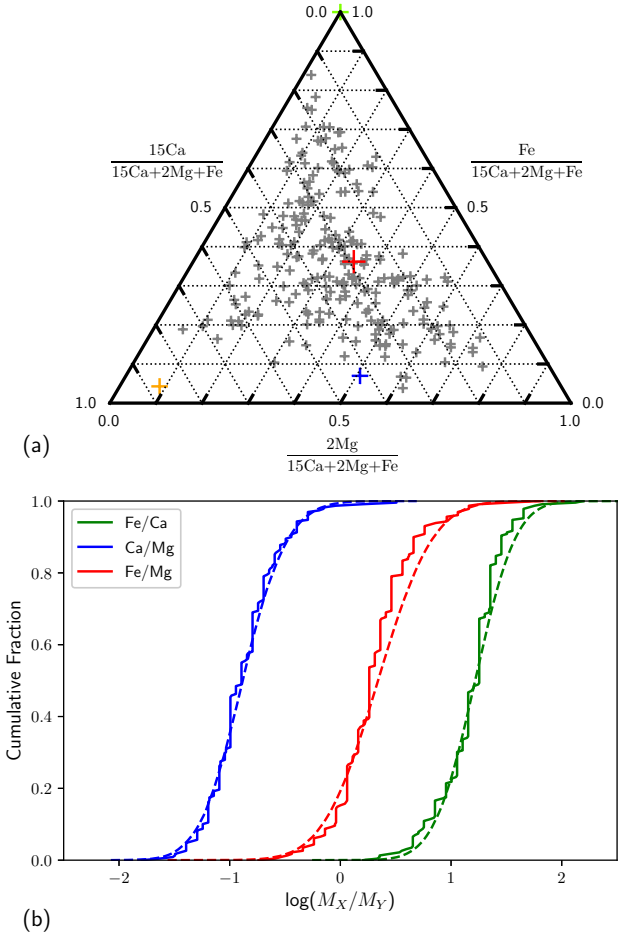}
    \caption{Model H - Best fit to the observations assuming that all the model compositions are given by single objects in the steady state phase.}
    \label{fig:SS}
\end{figure}

At this point, it is worth briefly looking at whether it is possible to find a good fit to the data under either the pre-steady state (Model G) or steady state (Model H) assumptions for single object accretion (see $\S$\ref{sec:StandAn}). The fit was performed over all 6 parameters as in $\S$\ref{sec:MgDep} but now the model predictions will simply be generated by picking objects randomly from the composition distribution. The minimum total values of $\chi^2$ for both models were found to be $92$ and $86$ for Models G and H respectively. The best fit parameters for both methods are also in Table \ref{tab:Parameters}. The best fit distributions are shown in Figures \ref{fig:pSS} and \ref{fig:SS} for the pre-steady state and steady state assumptions respectively.

From Table \ref{tab:Parameters} we can see that the two methods of considering single object accretion alone give somewhat different best fit values for the 6 parameters compared with full, stochastic model. Most notably, there is similarity in the value of $\mu_\text{Mg}$ for Models E and H and also for Models F and G. The reason for this is that Models F and G do not include any differential sinking whereas E and H do. It therefore makes sense that E and H require much larger Mg depletion to offset the effect of the differential sinking.

The ternary plots show one key difference when compared with the best fit for the full model in Figure \ref{fig:MgDep}. This is the lack of points in the bottom right corner of the plot. Under the full model, these points correspond to high Mg abundances caused by objects that have been in the atmosphere for several sinking times. Naturally, under the two single object assumptions, this cannot be recreated and so is absent from these plots. 

Overall, while these models give similar fits to the data as Model E, they are physically implausible, as we discuss at length in $\S$\ref{sec:LO}.

\section{Discussion}
\label{sec:Dis}

\subsection{To what Extent is Pollution Dominated by a Single Object?}
\label{sec:LO}

In $\S$\ref{sec:Results} we tested both our full model and the standard literature assumptions against the observed data. Here we consider for what fraction of observations the standard assumptions are expected to be valid on the assumption that the model in $\S$\ref{sec:model} is an accurate representation of the accretion process. The total mass of pollutant in the atmosphere, $m_\text{atm}$, is given by the sum of the $m_\text{atm,i}$ from each accretion event (see eq. \eqref{eq:sum}).

\begin{figure}
\includegraphics[width=\columnwidth]{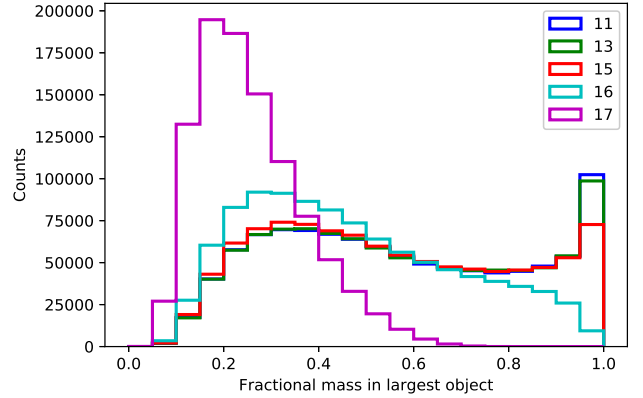}
\caption{Histograms showing the largest fraction of the pollutant mass in the atmosphere that arrived in the atmosphere in a single object. The plots are for different values of $\log(\dot{M}_\text{in}t_\text{samp})$ (in units of $\text{g}\,\text{s}^{-1}\,\text{yr}$). The height of the bars show the number of counts in each bin of width $0.05$.}
\label{fig:LO-tsink}
\end{figure}

Figure \ref{fig:LO-tsink} shows histograms for the largest value of $m_\text{atm,i}/m_\text{atm}$ (i.e. the fraction of the current total pollutant mass that originated in a single accretion event) at the time of observation. In general, this will depend both on $\dot{M}_\text{in}$ and $t_\text{samp}$. However, in the model, there is a degeneracy between $\dot{M}_\text{in}$ and $t_\text{samp}$ through the fact that the model output only depends on the combination $\dot{M}_\text{in}t_\text{samp}$, subject to a rescaling on the time and mass axes (which is not relevant here as we consider fractional masses).\footnote{In reality, $t_\text{samp}$ varies for different temperature stars and different elements. Because of this, we will here assume that all elements sink at the same rate and therefore ignore the composition of the accreted objects. The reason for this is that, if we were to include the different sinking times then we would need to choose a specific temperature for the white dwarf and therefore removing this will allow the conclusions to be applied more generally.}

Figure \ref{fig:LO-tsink} shows that for $\dot{M}_\text{in}t_\text{samp}=10^{13}\,\text{g}\,\text{s}^{-1}\,\text{yr}$, a single object contributes less than half of the observed mass $46\%$ of the time, and indeed that the assumption that the observed composition is dominated by a single object is invalid at least some of the time for all values of $t_\text{samp}$. There is also a clear trend that, as $\dot{M}_\text{in}t_\text{samp}$ increases, the likelihood that the observed composition is dominated by a single object decreases. This is due to the two regimes of continuous and stochastic accretion. If the accretion was purely stochastic then we would expect to see the vast majority of the counts in the $>0.95$ bin as it would be very rare to see multiple objects at once. However, if the accretion was purely continuous then we would expect to see very few counts in the region $\gtrsim 0.5$ as there would always be multiple bodies contributing significantly to the pollutant mass.

The majority of the histograms show this superposition between the continuous accretion (peaking around $0.3$) and the stochastic accretion (which gives the peak at $>0.95$). As $\dot{M}_\text{in}t_\text{samp}$ increases then the expected number of accretion events per sinking time also increases. Therefore, more of the mass distribution moves into the continuous regime and so the relative size of the peak at $0.3$ increases. There will therefore be a critical sampling time, $t_\text{samp,crit}$, above which all the accretion is approximately continuous. This is found by setting $m_\text{tr}=m_\text{max}$ in eq. \eqref{eq:mtr} (as in \citealt{2014MNRAS.439.3371W}) to give
\begin{equation}
t_\text{samp,crit} = m_\text{max}/\dot{M}_\text{in}.
\end{equation}
In these simulations this gives $\dot{M}_\text{in}t_\text{samp} = 3.2\times 10^{24}\,\text{g} = 1\times 10^{17}\,\text{g}\,\text{s}^{-1}\,\text{yr}$. Figure \ref{fig:LO-tsink} shows that, for this combination of sampling time and mass input rate, there are no counts in the purely stochastic regime and the plot is completely dominated by multiple accretion events, as expected.

Finally, we can see that the plots for $\log(\dot{M}_\text{in}t_\text{samp}) =$ 11 and 13 are virtually identical. For these values, the accretion of an object of $m_\text{max}$ is vanishingly rare. Therefore, since the shape of the mass distribution is constant and extends to arbitrarily small masses, even smaller values of $\dot{M}_\text{in}t_\text{samp}$ would not result in any change in the shape of the plot. Therefore, we expect accretion of multiple objects to be important in a large proportion of situations although accretion of single objects can contribute a significant fraction of the pollutants some of the time.

To this point, this analysis has assumed that all possible pollutant levels are equally likely to be observed. However, we showed in $\S$\ref{sec:thresh} that in reality the pollution is only observable for half of the time. We therefore crudely remove the data with the lowest total mass of pollutant. This is a proxy for the fact that stars whose pollutants are below a certain threshold will not be present in the sample as their metal lines will be below the level required for detection.

\begin{figure}
\includegraphics[width=\columnwidth]{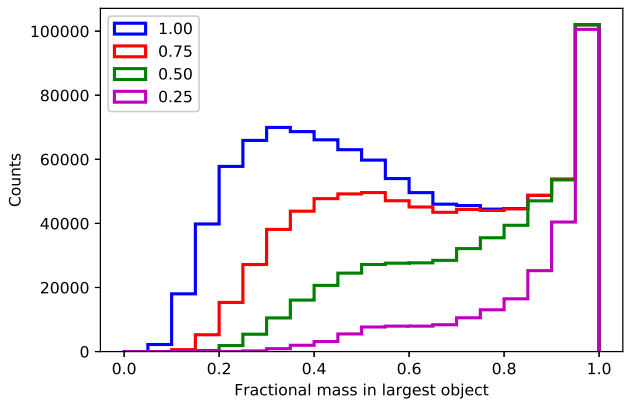}
\caption{Histograms showing the largest fraction of the pollutant mass in the atmosphere that arrived in the atmosphere in a single object. The plots all have $\dot{M}_\text{in}t_\text{samp}=10^{13}\,\text{g}\,\text{s}^{-1}\,\text{yr}$. The different plots are created with different fractions of the data (as shown in the legend). In all cases, the data with the lowest total mass of pollutants has been discarded.}
\label{fig:LO-thresh}
\end{figure}

The result of removing the weak data is shown in Figure \ref{fig:LO-thresh}. The plots show the effect of removing various fractions (as indicated in the legend) of the data before performing the analysis. We can see that removing the data with the lowest total mass of pollutant preferentially removes the observations which are dominated by multiple bodies. We can understand this by looking at Figure \ref{fig:Mod2}. This plot shows that the samples where the total mass is large are those immediately following the accretion of a massive body. The samples with a small total mass (and therefore those that are preferentially removed) are those where there has been no recent accretion of a massive body and therefore the pollutant levels are in the continuous regime where there are multiple contributing objects.

With all the data, the majority of the total pollutant mass can be attributed to a single body (i.e. $>0.5$ on the histogram) $54\%$ of the time. When half of the data is removed (similar to the value found in $\S$\ref{sec:thresh}), that percentage drops to $21\%$. While $21\%$ might seem like a relatively low percentage, it is important to remember that even when a single body is contributing $\sim 80\%$ of the mass, the other $\sim 20\%$ still affects the analysis in a way that can be quantified in the context of the model presented in $\S$\ref{sec:model}.

The overall implication of introducing an observational threshold is that the higher the threshold, the more accurate the assumption that a single object dominates the observed pollutant composition at any given time. However, for the sample considered in this paper, we have shown that it is essential to consider multiple accretion events in order to gain a thorough understanding of the composition of the accreted planetesimals.

We are therefore able to say that, for the sample and model considered in this paper, Models G and H (which assume the observed composition originates from a single object and are standard models used in the literature) are implausible as mentioned earlier. One potential way in which Models G and H might be plausible would be if the planetesimals are all of the same size. Therefore, if the accretion rate is small enough, the accretion would be stochastic with minimal overlap in pollution originating from different bodies. However, this was found to be incompatible with the observed data by \citet{2014MNRAS.439.3371W}. Furthermore, it is true that a mass distribution (as used here) is a more realistic model, as can be seen by comparison with our own asteroid belt.

The consequences of this multi-body accretion are primarily twofold. Firstly, this result implies that white dwarfs which appear to be polluted by material of a primitive composition may instead have been polluted by multiple differentiated bodies whose compositions average to primitive. This means that any analysis that assumes the pollution is from a single body will find too many primitive bodies.

The second consequence is that we can now provide a mechanism whereby the observed composition could be composed of core and crust, while poor in mantle. This composition is not predicted by collisional models \citep{2018E&PSL.484..276C} and as such, when considering single object accretions only, it is rejected as a potential explanation (e.g. \citealt{2018MNRAS.479.3814H}) even when it provides statistically the best fit. With multiple accretion events, it is very plausible that a white dwarf could accrete two planetesimals, one core-rich and the other crust-rich, which together appear to be a mixture of core and crustal material.

\subsection{Implications of Compositional Fit}
\label{sec:CompFit}

\subsubsection{Variation of observation versus that of stellar material}

The large difference in the variation between the observed composition (as shown in Figure \ref{fig:Obs-Norm}) and the composition of the sample of FGK stars (Figure \ref{fig:Obs-Stel}) shows that stellar variation cannot explain the variation in the observed composition. This is therefore further evidence that the source of the atmospheric pollution cannot be stellar material \citep{2014AREPS..42...45J}.

\subsubsection{Mg depletion}

The most obvious result from trying to fit to the observed composition is the apparent lack of Mg in the observations. The mean of the sample is found close to the bulk Earth composition \citep{2018MNRAS.477...93H} (which in turn lies within the distribution of the FGK stars). However, since Mg sinks much more slowly than either Ca or Fe, we would expect it to be enhanced in the observations. We therefore present three potential explanations.

\begin{enumerate}
\item The first potential explanation is that the observed Mg depletion is a real effect. The enhancement or depletion in certain elements is not a new effect (e.g. \citealt{2012ApJ...749....6D}). In a protoplanetary disc, each element has its own ice line. Interior to this line the element is in its gaseous form and so is not expected to be present in planetesimals formed in this region. The location of the ice lines are determined by each element's position in the refractory series with elements higher in the series having higher sublimation points and so ice lines that are closer to the parent star. The condensation temperatures in this case are $T_\text{Ca}=1659\,\text{K}$, $T_\text{Mg}=1397\,\text{K}$ and $T_\text{Fe}=1357\,\text{K}$ \citep{2003ApJ...591.1220L}. For Mg depletion to occur during formation, $T>1397\,\text{K}$ is required. However, this would also produce Fe depletion which we do not see in this case.

One potential explanation is that the depletion occurred not at formation, but during the giant branch. During the giant branch, planetesimals close enough to the star will be heated to temperatures that may exceed the sublimation temperature, and this may be different for minerals that contain different abundances of Mg. To deplete the Mg, we expect that material from the mantle would need to be preferentially removed. The plausibility of this explanation requires a consideration of material physics which is beyond the scope of this paper.

Observations of $\beta$ Pictoris have shown the presense of extremely Mg-rich olivine crystals \citep{2012Natur.490...74D}. Although this observation shows Mg enhancement rather than depletion (as required for our model) it is interesting to note that processes like this can occur.

\item If instead of the sinking times given in \citet{2017MNRAS.467.4970H} all the elements (Ca, Mg and Fe) are depleted at the same rate then there would be no differential sinking. This was explored in $\S$\ref{sec:Sink} where it was found that with identical sinking times the best fit value of $\left\langle f_\text{Dep,Mg}\right\rangle$ increased from 0.22 (with differential sinking) to 1.03 (without differential sinking). This drastic increase was found with only a crude implementation of identical sinking times for all elements and suggests that adjustment to the sinking times used in \citet{2017MNRAS.467.4970H} could remove a large fraction of the required Mg depletion. 

One possible mechanism whereby this could occur is thermohaline convection. Thermohaline instabilities result from an inverted mean molecular weight gradient in the atmosphere of white dwarfs. This instability results in the depletion of all elements equally \citep{2018ApJ...859L..19B}. However, thermohaline instabilities only dominate over gravitational settling if they occur on a shorter timescale, and \citet{2018ApJ...859L..19B} conclude that thermohaline instability only dominates in white dwarfs with thin, hot, H dominated atmospheres. The sample we have used consists of cool white dwarfs with He dominated atmospheres which are expected to be dominated by gravitational settling. However, if this is incorrect then this could be a potential explanation.

Alternatively, it could be that gravitational settling is the dominant mechanism but that the current methods for calculating the timescales for different elements are incorrect. It is notable that the timescales given in \citet{2009A&A...498..517K} show a much smaller difference between the sinking timescale for Mg and those for Ca and Fe than those given in \citet{2017MNRAS.467.4970H}.

\item The final explanation is that the lack of Mg is due to biases in the data. In $\S$\ref{sec:thresh} we implemented a crude threshold for detection into our model. Although this had only a minor effect on the issue of Mg depletion this is only one possible way in which biases could have entered the sample. In order to select the sample in \citet{2017MNRAS.467.4970H} it was necessary to perform cuts in the data to reduce the sample size for visual inspection. It is therefore possible that these cuts preferentially removed stars with high levels of Mg (e.g. if the levels of pollutant alters the position on the colour plot) which could lead to the apparent lack of Mg in the sample.
\end{enumerate}

\subsubsection{Core, mantle and crust fractions}

\begin{table}
    \centering
    \caption{Comparison of core, mantle and crust fractions for Earth and our best fit model (Model E). The values for Earth are those given in eq. \eqref{eq:fractions}.}
    \label{tab:fractions}
    \begin{tabular}{lccc}
        \hline
         & $f_\text{Cru}$ & $f_\text{Man}$ & $f_\text{Cor}$ \\
        \hline
        Earth & 0.000148 & 0.675 & 0.325 \\
        Model &0.15 & 0.49 & 0.36 \\
        \hline
    \end{tabular}
\end{table}

\begin{figure}
	\includegraphics[width=\columnwidth]{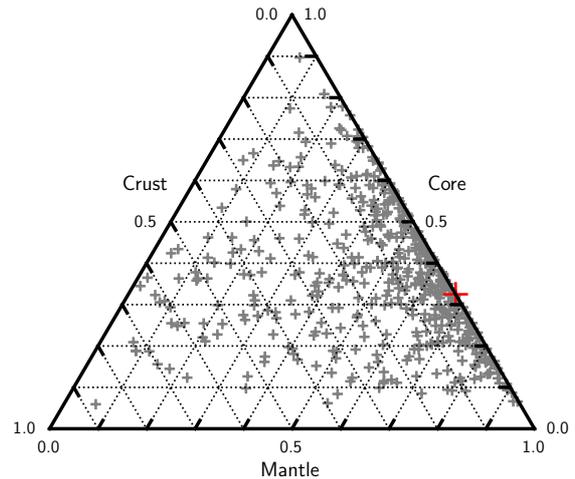}
    \caption{The distribution of the crust, mantle and core fractions of 500 randomly chosen planetesimals taken from the distributions using the best fit parameters for Model E (given in Table \ref{tab:Parameters}). Also shown is the position of the Earth (red).}
    \label{fig:Final}
\end{figure}

Table \ref{tab:fractions} gives the core, mantle and crust fractions for the Earth and for the best fit of our full model (Model E). Figure \ref{fig:Final} shows the distribution of the crust, mantle and core fractions of 500 randomly chosen planetesimals given our best fit parameters for Model E (given in Table \ref{tab:Parameters}).

In Figure \ref{fig:Final}, $42\%$ of the planetesimals have large ($>0.1$) crust fractions. In comparison, the Earth's crust fraction is on the order of $10^{-4}$ and so our model has a large number of objects with crust fractions $\sim3$ orders of magnitude larger than this. While there is no reason as to why the Earth should be typical of planetesimals around white dwarfs, it is nonetheless an interesting comparison.

\subsection{What happens when Sinking Timescales are much shorter than the Disc Timescale?}
\label{sec:Disc}

As mentioned earlier, the model (specifically the combination of $t_\text{sink}$ and $t_\text{disc}$ into $t_\text{samp}$) breaks down if $t_\text{disc}\gtrsim t_\text{sink}$. To enable us to make predictions for white dwarfs with small sinking times we therefore need to adjust our treatment of the disc. To do this, we need to consider the disc and the atmosphere separately. In this section we consider two related but distinct models for the disc.

Firstly, in \citet{2009A&A...498..517K} it is assumed that the rate of accretion from the disc into the atmosphere is constant over the disc timescale with an instantaneous start and end to the accretion. This is necessary for the three phases for a single object as in $\S$\ref{sec:StandAn}. In our model with a timestep of $\text{d}t$ the accretion from a single object onto the atmosphere lasts for $t_\text{disc}/\text{d}t$. This is implemented by replacing the instantaneous deposition of the entire mass onto the atmosphere with a deposition that is spread out over the next $(t_\text{disc}/\text{d}t+1)$ timesteps ($+1$ to include the current timestep when the object enters the disc). Thus, if $t_\text{sink}\gg t_\text{disc}$ the mass will be spread out only over the current timestep and so the original model is recovered, whereas for disc timescales that are larger than the sinking time, the mass will be spread out over many timesteps. This model will be referred to as the constant disc model.

The second disc model assumes the mass in the disc decays exponentially (ignoring any material added to the disc from new objects) on the disc timescale and deposits the lost mass onto the atmosphere. This model fits with the procedure used to create $t_\text{samp}$ in eq. \eqref{eq:tsamp}. In the limit that $t_\text{disc}\gg t_\text{sink}$, eq. \eqref{eq:tsamp} becomes $t_\text{samp}\sim t_\text{disc}$. Since under the full model the mass decays exponentially on this sampling time, the original model also treats the disc exponentially (as the disc timescale becomes an exponential timescale under certain conditions). This is implemented in the model simply by separately tracking the disc and atmosphere pollutant mass. Each new object enters the disc which in turn feeds the atmosphere. If $t_\text{sink}\gg t_\text{disc}$ then $\text{d}t\gg t_\text{disc}$ as well and so the entire mass of the disc is deposited onto the atmosphere effectively instantaneously, again recovering the original model. This model will be referred to as the exponential disc model.

\begin{figure}
	\includegraphics[width=\columnwidth]{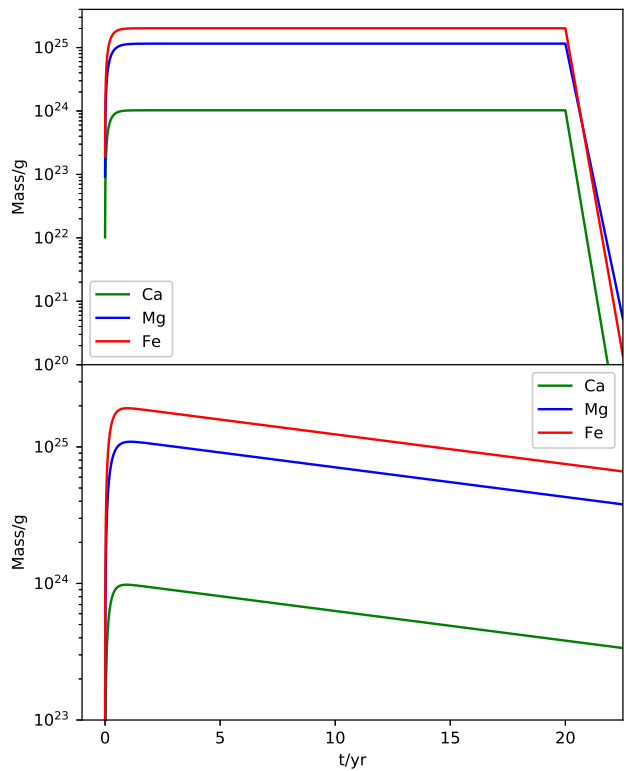}
    \caption{Time variation of the pollutant mass remaining in the atmosphere of G$\, 29$-$38$, assuming that it accreted an object with a composition equivalent to that of the Earth. The two panels show the results under the constant disc model (top) and the exponential disc model (bottom).}
    \label{fig:Disc1}
\end{figure}

In our analysis of the effect of this updated model we will use sinking times for G$\, 29$-$38$. These times are $0.20$, $0.25$ and $0.21\,\text{yr}$ for Ca, Mg and Fe respectively \citep{2014ApJ...783...79X}. Before we explore the effect of the disc on the full model, Figure \ref{fig:Disc1} shows the effect of the disc for a single object accretion. The sinking timescale is the rough timescale on which the levels of atmospheric pollution can adjust to a new steady state. For G$\, 29$-$38$, the sinking times are around $100$ times smaller than the disc timescale and so Figure \ref{fig:Disc1} shows that the pollutant levels build up to these steady states very quickly for both disc models.

The constant disc exhibits the same behaviour as that given in \citet{2009A&A...498..517K} and summarised in $\S$\ref{sec:StandAn}. During the steady state phase the pollutant levels are constant and then decay exponentially when accretion from the disc ceases. The pollutant levels drop according to the sinking timescale of each element and so the relative enrichment of Mg is again seen in this declining phase.

For the exponential disc, accretion from the disc onto the atmosphere does not cease. Since the sinking timescales are much shorter than the disc timescale, the atmosphere pollutant levels are able to adjust quickly to the decreasing accretion level from the disc. For this reason, the levels of each element decay at the same rate (the disc timescale) and so the composition will always appear as if it is in the pre-steady state or steady state phase and never enter the declining phase.

\begin{figure}
	\includegraphics[width=\columnwidth]{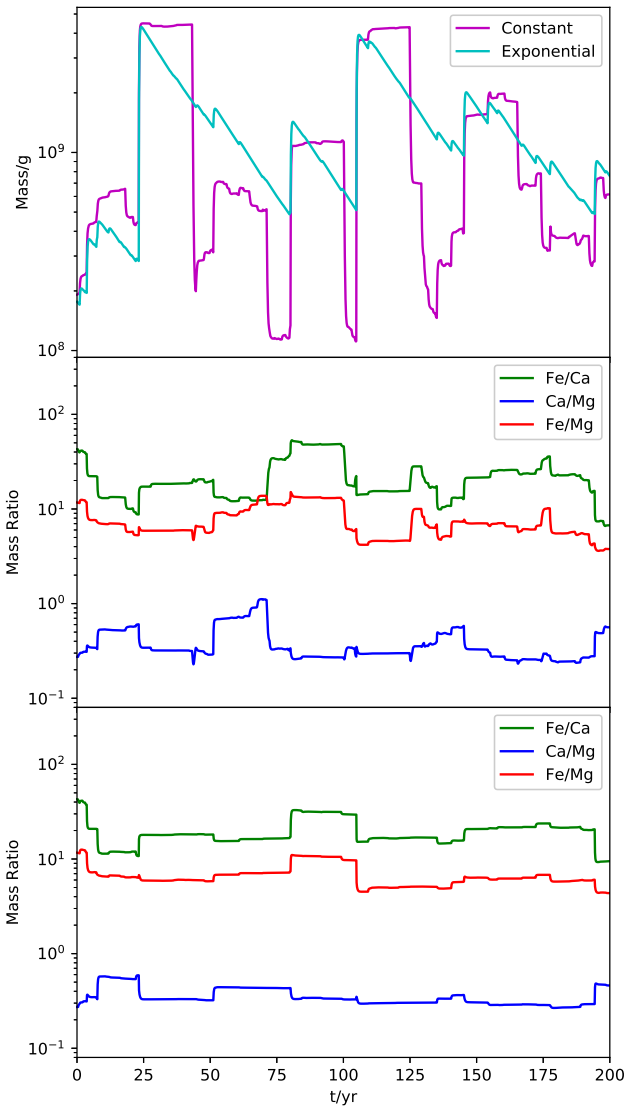}
    \caption{The pollutant mass and compositional variation with time under the two disc models. The top panel shows the total mass of Ca, Mg and Fe in the atmosphere under each model. The middle panel shows the three mass ratios under the constant disc model and the bottom panel the same for the exponential disc model. The simulation is performed with the parameters in Table \ref{tab:Parameters} with the exception that $\dot{M}_\text{in}=10^8\,\text{g}\,\text{s}^{-1}$ and Mg depletion was chosen to be the mean value of $\left\langle f_\text{Dep,Mg}\right\rangle=0.22$.}
    \label{fig:Disc2}
\end{figure}

Figure \ref{fig:Disc2} shows the result of using the full, stochastic model with both of the disc implementations. For each disc, the same inputs (i.e. sizes and times of asteroid accretion) were used to allow direct comparison. The top panel shows that the two discs give similar levels of pollutant mass in the atmosphere. However, it is clear from this that high pollutant levels persist for much longer after the accretion of a massive object under the exponential model than under the constant model. This is in line with what is expected given the results for single objects in Figure \ref{fig:Disc1}.

The plots of the mass ratios in Figure \ref{fig:Disc2} show that a greater degree of variation is seen under the constant disc than the exponential disc. This is because under the constant disc model, a single object can only dominate the composition for $20\,\text{yr}$ (the disc timescale) whereas under the exponential disc it can dominate for much longer. However, under both disc models variation in the observed composition occurs on a timescale similar to the disc timescale.

Under both disc models the variation of the observed composition occurs on a timescale of $\sim 20\,\text{yr}$ (the value depending somewhat on the disc model). Therefore, to a first approximation it varies on the disc timescale. This is in contrast to the situation when $t_\text{disc}\ll t_\text{sink}$. In this case the disc timescale is so short that the disc makes a negligible difference to what is observed and the pollutant material for any given planetesimal effectively arrives in the atmosphere as one. In this case, a first approximation is that the composition would be expected to vary on the sinking timescale as this is now the timescale over which any given object can dominate.  Therefore, in general we expect the observed composition to vary on the longer of the sinking and disc timescales.

\subsection{Caveats}
\label{sec:Cav}

There are a number of assumptions that have been made in this paper, the majority of which are for the sake of the simplicity of the model. Many of these assumptions are the same as those in \citet{2014MNRAS.439.3371W} as this model is an extension of theirs. Therefore, the caveats mentioned there are also applicable here.

In the original model in \citet{2014MNRAS.439.3371W} the only element considered was Ca. The total mass of pollutant was then approximated by assuming that the Ca fraction in the polluting body was the same as the Ca fraction of the bulk Earth. We have shown that a variation in the Ca fraction of polluting bodies is required to explain the observations. However, this variation is much less than that in the total mass of pollutant (which spans several orders of magnitude). Therefore, while the true mass of pollutant will show more variation than that assumed in \citet{2014MNRAS.439.3371W} this is likely to be only a very small correction.

Although we argued earlier that thermohaline instabilities do not affect the stars in this sample, it would affect some of the sample used in \citet{2014MNRAS.439.3371W}. It would preferentially affect the hot stars with short gravitational settling times and also affect stars with greater values of $\dot{M}_\text{in}$ \citep{2018ApJ...859L..19B}. This would reduce the sinking time of these stars and so the observed values of $\dot{M}_\text{in}$ would be greater. Although a full understanding of the consequences would require the analysis from \citet{2014MNRAS.439.3371W} to be adjusted (which may not be straightforward) we can still qualitatively consider its repercussions.
\begin{itemize}
\item The value of $m_\text{max}$ was only constrained by a lower limit (above this its value had a negligible effect) and so any changes in the inferred $\dot{M}_\text{in}$ due to changes to $t_\text{sink}$ are unlikely to change the value of $m_\text{max}$ drastically.
\item The parameters from $\dot{M}_\text{in}$ ($\mu$ and $\sigma$) were found by fitting to the longest sinking time bin. As this is expected to be relatively unchanged from thermohaline considerations it is unlikely that the parameters will change either.
\item The final two parameters, $q$ and $t_\text{disc}$ were found by fitting to the short and medium sinking time bins. The short bin will certainly be changed by the inclusion of thermohaline instabilities while the medium bin is likely to be changed but to a much lesser extent. Without performing the full fit, it is difficult to be certain as to what effect this makes. One possible effect is that $t_\text{disc}$ would need to increase to provide the increase in the inferred $\dot{M}_\text{in}$ for the stars with short sinking times.
\end{itemize}

Therefore, it is likely that the most significant changes to the values in Table \ref{tab:2014-values} are to the values of $q$ and $t_\text{disc}$. Any increase to the disc timescale may go some way to alleviating the tension between the disc timescale of $20\,\text{yr}$ as found in \citet{2014MNRAS.439.3371W} and the values given in the majority of the literature (e.g. \citealt{2012ApJ...749..154G}). \citet{2012ApJ...749..154G} give a value on disc lifetimes of $\log\left(t_\text{disc}/\text{yr}\right)=5.6\pm 1.1$. However, this is the time from the initial formation of the disc to the point where it has disappeared. The important timescale in this model is the time from when the first part of an object arrives on the white dwarf to when the last part does. Since we would expect the disc to form before the first part of the object reaches that atmosphere of the white dwarf, this value is likely to be larger than the value required by the model.

With that in mind, the value of $\log(t_\text{disc})=5.6\pm 1.1$ is within $1\sigma$ of the average of the sinking times of our sample of DZ white dwarfs. With a disc timescale of only $20\,\text{yr}$, it was impossible for any accretion to reach a steady state phase. With a much longer disc timescale, it may be possible for individual objects to be in the steady state phase. However, we would still expect multiple objects to be contributing to and for differential sinking to be important in the observed composition. We therefore do not expect this to have a large effect on our conclusions.

Of perhaps more relevance is any change to the value of $q$ which is responsible for the relative contributions of the high and low mass ends of the mass distribution to the total mass budget. A large value of $q$ corresponds to a steep distribution with very few large objects whereas a small value of $q$ would imply comparatively more large objects. This could lead to changes to the histograms in Figure \ref{fig:LO-tsink} where large values of $q$ would lead to a reduction in the peak at $>0.95$ (which corresponds to the recent accretion of a single large object) compared to that near $0.3$ and vice versa for a small value of $q$. Therefore, with a value of $q<1.57$, the observed composition would be dominated by a single object more often, and Models G and H (the standard literature models) would be more accurate than they are with $q=1.57$ (the value used in our model).

In the parameterisation of the composition of the accreted planetesimals it has been necessary to make many simplifying assumptions. Without these assumptions, the number of free parameters in the model would have become too large to manage and it would have been impossible to constrain the output values from the model. Despite this, it is worth noting that the logit-normal and log-normal distributions for the free parameters have been chosen for simplicity rather than for any physical reason. It is very plausible that (for example) the true mantle fraction does not follow a simple distribution but is an altogether more complicated function. However, introducing further complexity would likely lead to degeneracy in the final best fit values.

As we mentioned in $\S$\ref{sec:Samp} and $\S$\ref{sec:Stel}, the observed data have errors which we have not included in our analysis. These errors were given as lying in the range of $0.05-0.3$ dex. The effect of these errors would be to provide some of the spread seen in the observations. We therefore expect that the true width of the composition distributions are narrower than we have found here. It is important to note that the errors would not affect the mean of any of our distributions. Importantly, this means that the Mg depletion which was required would not be affected by the errors.

For both of the above reasons, the values concluded in Table \ref{tab:Parameters} should therefore not be treated as predictions of the true physical distributions but simply as the values which provide the best fit under the assumptions we have outlined. However, despite this caveat, we are still able to draw conclusions from the general trends which our final values indicate.

\section{Conclusions}
\label{sec:Conc}

In this paper we have extended the model presented in \citet{2014MNRAS.439.3371W} to include consideration of the composition of the accreted planetesimals. In $\S$\ref{sec:Results} we first used this model with an increasingly complex parameterisation for the composition of the accreted planetesimals to try and produce a good fit to the observations. It was found that a simple assumption that each body is formed only from core or mantle-like material with the fractions of each body given by a Gaussian parameterisation could produce a good fit to the distribution of the Fe/Ca ratio. However, this produced a very poor fit to the Fe/Mg and Ca/Mg distributions showing that the use of multiple species can give much better constraints on pollutant compositions.

We then relaxed the restriction on only core and mantle-like material to allow for crust-like material. However, this had only a minor effect on the fit. The main issue was an apparent systematic offset in the Fe/Mg and Ca/Mg distributions between the model output and the observations. This lead to the introduction of a parameterisation for the Mg depletion of the accreted bodies, which removed the systematic offset. With these 6 parameters (from two logit-normal distributions for the mantle and crust fractions and a log-normal distribution for the Mg depletion), it was possible to produce a good fit to all three observed distributions subject to a few deviations.

In $\S$\ref{sec:Results} we also tested models with adjusted sinking times (such that all elements sank on the same timescale) and models based off standard literature assumptions of single objects being accreted in pre-steady and steady state. While all three of these new models gave good fits to the data, the parameters required to give these fits varied widely.

In $\S$\ref{sec:LO} we then tested the standard assumption that the observed pollution is dominated by a single object. It was found that, while this assumption holds true for a reasonable proportion of the time, it is not universally applicable and a minimum estimate is that it is incorrect $20\%$ of the time. As it is impossible to know for which observations the assumption is valid, we can conclude that consideration of multiple accreting bodies is essential to gain a thorough understanding of the composition of the planetesimals responsible for the pollution of white dwarfs.

The implications of the best fit parameters from the complete model were considered in $\S$\ref{sec:CompFit}. The values were compared to the bulk Earth values and it was found that our model requires a much larger crust fraction (by $\sim3$ orders of magnitude) than that of the Earth. Perhaps the most surprising conclusion was that we required $\left\langle f_\text{Dep,Mg}\right\rangle=0.22$ to provide a good fit to that data. This corresponds to the accreted planetesimals having (on average) only $22\%$ of the Mg levels of their parent star (compared to Ca and Fe). While this could be due to a true depletion in Mg it is also possible that it originates from errors in the sinking times used in the model or from a potential bias in composition in the observed sample. 

In $\S$\ref{sec:Disc} we were then able to use our model to predict the time variation of the observed composition of any one white dwarf. It is found that the variation occurs on a timescale similar to the longer of the sinking and disc timescales. This could therefore be used in the future to infer lower limits on disc timescales by observing the change in pollutant composition over time.

The conclusions presented here come from a relatively simple model for planetesimal composition and have been fit to the first (potentially biased) large sample of its kind. There are therefore two separate avenues in which further work could be performed. The first is in including more accurate models for the planetesimal composition (e.g. \citealt{2018MNRAS.479.3814H}) into our Monte Carlo model. However, as mentioned before, doing this would introduce many new degrees of freedom which would likely lead to degeneracy in the output values. The second avenue is to apply this model to other surveys of polluted white dwarfs. The survey by \citet{2017MNRAS.467.4970H} is currently the only survey of its kind. However, new surveys such as Gaia could potentially provide large numbers of hitherto unknown polluted white dwarfs on which this model could be further tested.

\section*{Acknowledgements}

The authors would like to thank the anonymous referee whose review improved the quality and clarity of this work. SGDT would like to thank Amy Bonsor and John Harrison for useful discussions on the underlying physics. SGDT acknowledges funding received from a Royal Astronomical Society Undergraduate Summer Bursary.




\bibliographystyle{mnras}
\bibliography{references} 






\bsp	
\label{lastpage}
\end{document}